\documentclass[preprint,12pt]{elsarticle}

\usepackage{amssymb}

\usepackage{amsmath}

\newcounter{bla}

\usepackage{subcaption}
\usepackage{listings}
\usepackage{xcolor}
	
\usepackage{soul}

\definecolor{codegreen}{rgb}{0,0.6,0}
\definecolor{codegray}{rgb}{0.5,0.5,0.5}
\definecolor{codepurple}{rgb}{0.58,0,0.82}
\definecolor{backcolour}{rgb}{0.95,0.95,0.92}

\lstdefinestyle{mystyle}{
    backgroundcolor=\color{backcolour},   
    commentstyle=\color{codegreen},
    keywordstyle=\color{magenta},
    numberstyle=\tiny\color{codegray},
    stringstyle=\color{codepurple},
    basicstyle=\ttfamily\footnotesize,
    breakatwhitespace=false,         
    breaklines=true,                 
    captionpos=b,                    
    keepspaces=true,                 
    numbers=left,                    
    numbersep=5pt,                  
    showspaces=false,                
    showstringspaces=false,
    showtabs=false,                  
    tabsize=2
}
\colorlet{punct}{red!60!black}
\definecolor{background}{HTML}{EEEEEE}
\definecolor{delim}{RGB}{20,105,176}

\lstdefinelanguage{json}{
    basicstyle=\normalfont\ttfamily,
    numbers=left,
    numberstyle=\scriptsize,
    stepnumber=1,
    numbersep=8pt,
    showstringspaces=false,
    breaklines=true,
    frame=lines,
    backgroundcolor=\color{background},
    literate=
     *{:}{{{\color{punct}{:}}}}{1}
      {,}{{{\color{punct}{,}}}}{1}
      {\{}{{{\color{delim}{\{}}}}{1}
      {\}}{{{\color{delim}{\}}}}}{1}
      {[}{{{\color{delim}{[}}}}{1}
      {]}{{{\color{delim}{]}}}}{1},
}
\usepackage[outdir=./]{epstopdf}

\usepackage{booktabs}

\usepackage{tabularx} 
\journal{Computer Physics Communications}

\begin{document}

\begin{frontmatter}



\title{ReMKiT1D - A framework for building reactive multi-fluid models of the tokamak Scrape-Off Layer with coupled electron kinetics in 1D}


\author[a]{Stefan Mijin\corref{author}}
\author[b]{Dominic Power}
\author[a,c]{Ryan Holden}
\author[a]{William Hornsby}
\author[a]{David Moulton}
\author[a]{Fulvio Militello}

\cortext[author] {Corresponding author.\\\textit{E-mail address:} stefan.mijin@ukaea.uk}
\address[a]{United Kingdom Atomic Energy Authority, Culham
Campus, Abingdon, Oxfordshire, OX14 3DB, UK}
\address[b]{Blackett Lab., Plasma Physics Group, Imperial College London,
London, SW7 2AZ, UK}
\address[c]{School of Mathematics and Physics, University of Surrey, Guildford, GU2 7XH, UK}

\begin{abstract}

In this manuscript we present the recently developed flexible framework for building both fluid and electron kinetic models of the tokamak Scrape-Off Layer in 1D - ReMKiT1D (\textbf{Re}active \textbf{M}ulti-fluid and \textbf{Ki}netic \textbf{T}ransport in \textbf{1D}). The framework can handle systems of non-linear ODEs, various 1D PDEs arising in fluid modelling, as well as PDEs arising from the treatment of the electron kinetic equation. As such, the framework allows for flexibility in fluid models of the Scrape-Off Layer while allowing the easy addition of kinetic electron effects. We focus on presenting both the high-level design decisions that allow for model flexibility, as well as the most important implementation aspects. A significant number of verification and performance tests are presented, as well as a step-by-step walkthrough of a simple example for setting up models using the Python interface. 

\end{abstract}


\end{frontmatter}



{\bf PROGRAM SUMMARY}

\begin{small}
\noindent
{\em Program Title: ReMKiT1D}                                          \\
{\em CPC Library link to program files:} (to be added by Technical Editor) \\
{\em Developer's repository link:} https://github.com/ukaea/ReMKiT1D and \\ https://github.com/ukaea/ReMKiT1D-Python \\
{\em Code Ocean capsule:} (to be added by Technical Editor)\\
{\em Licensing provisions:} GPLv3 \\
{\em Programming language:} Fortran, Python                                  \\
{\em Supplementary material: https://doi.org/10.14468/fdq7-z869}                                 \\
{\em Nature of problem:} The flexible generation and modification of 1D models pertaining to multi-fluid simulations of the tokamak Scrape-Off Layer (SOL) with electron kinetics and reaction support. This would then allow both for rapid iteration on reduced models as well as the evaluation of kinetic electron effects in equilibria and during transients, following the formalism previously developed for SOL-KiT[1]. The framework was not only envisioned as the successor to SOL-KiT, but a tool that would allow users to construct their own models coupled with electron kinetics capabilities.\\
{\em Solution method:} The framework is written heavily utilizing Object-Oriented design principles, in particular using an extended version of the puppeteer pattern as presented by Rouson et al[2], as well as the heavy use of the strategy pattern/dependency injection. The Fortran code is MPI parallel and utilizes the PETSc library for implicit time-stepping. MPI parallelization is extended to distribution function Legendre harmonics, allowing for improved strong scaling. Initialization of the Fortran framework is done using JSON configuration files generated by an accompanying Python interface, and data analysis is standardized using widely used data formats such as HDF5.\\
{\em Additional comments including restrictions and unusual features:} The present manuscript focuses on the design and high-level implementation of the framework, as well as the demonstration of the workflow and various verification and performance benchmarking tests. Some details are avoided for the sake of brevity at various points, and these are meant to be available as part of the general code documentation or tutorials offered on the main repositories.\\

\end{small}

\section{Introduction}
\label{Introduction}

The Scrape-Off Layer (SOL) denotes the region of open magnetic field lines just outside of the core of magnetically confined fusion (MCF) devices, such as the tokamak. These open field lines impinge on material surfaces (walls, limiters, or divertor targets) in the device, leading to plasma-surface interactions, which can inject impurity species into the plasma. Furthermore, the region outside the core, often referred to as the edge, is cold enough for atomic, or even molecular, neutrals to persist and interact with the plasma in a non-trivial way. As such, this region of the device is particularly rich in multi-species plasma physics, with many reactions occurring between the species. 

Furthermore, due to the scale separation of parallel (to the open magnetic field lines) and perpendicular transport, 1D analytical and numerical models of the SOL have been a mainstay since the early days of SOL simulations. The reduced geometrical/dimensional complexity allows for the treatment of more involved physics, while keeping run times short compared to 2D or 3D simulations. As such, a range of 1D codes geared towards exploring one or more aspects of the SOL have been in use over the last several decades, ranging from PIC codes\cite{Tskhakaya2012,Procassini1992}, to continuum fluid\cite{Dudson2019,Derks2022,Havlickova2013} and kinetic \cite{Batishchev1999,Chankin2014,Zhao2019,Kupfer1996} codes. These have been used for a variety of problems, and some of them will be reviewed here to provide context for the code to be presented in the bulk of this text. 

One of the most common problems tackled by 1D codes is that of simulating equilibrium divertor regimes, with a particular focus on detachment\cite{Krasheninnikov2017,Stangeby2018,Dudson2019} - the regime in which the plasma recombines significantly in front of the target, effectively producing a cloud of neutrals which shields it from exposure to the hot upstream conditions. Tying into this is the study of how target conditions are affected by transient phenomena upstream\cite{Tskhakaya2008a,Vasileska2021}, when particles and energy are transiently injected far away from the targets and allowed to propagate towards them and interact with any of the background plasma and neutral species present in the equilibrium conditions. 

Another field of research is that of parallel transport in the SOL. While fluid models tend to use classical values of transport coefficients (e.g. heat conductivity or viscosity) that rely on local values of plasma parameters, due to the strong parallel gradients formed in the SOL, kinetic effects can come into play\cite{Fundamenski2005}, modifying the classical values in both equilibrium and transient conditions. Examples of kinetic effects include heat flux suppression and enhancement\cite{Brodrick2017,Chankin2018,Power2021}, as well as the modification of the target plasma sheath properties\cite{Mijin2020a,Power2023}, which fluid codes struggle to include consistently, often resorting to measures such as heat flux limiting. 

Finally, with the presence of many species, and atomic and molecular physics coming into play, the need arises to couple the various species through reaction rates, leading naturally towards collisional-radiative models (CRMs), which attempt to solve and reduce a set of coupled ODEs\cite{Bates62,Bates1964,Sawada1995,Summers2006,Wunderlich2016,Greenland2001}. Most often, however, these models are 0D, neglecting transport effects and taking reaction rates based on an underlying Maxwellian distribution for the plasma species. 1D models, especially kinetic models, when including collisional-radiative processes, can be used to study transport effects on the particle and energy balance governed by these complicated reactions\cite{Mijin2020b}. 

The goal of the framework to be presented is to provide a relatively easy way to build models that can tackle most of the above issues, combining multi-fluid, collisional-radiative, and kinetic physics, particularly in the context of SOL plasmas. However, the presented framework could also be relevant to other fields where 1D fluid equations might need to be solved. 

The ReMKiT1D (\textbf{Re}active \textbf{M}ulti-fluid and \textbf{Ki}netic \textbf{T}ransport in \textbf{1D}) framework consists of a core code written in Modern Fortran and controlled through JSON configuration files constructed using a Python package, allowing for rapid model iteration and flexibility. While the numerical approach is extremely customizable, with many features exposed to the user at the Python level, a significant number of numerical procedures and approaches are adapted from work previously done on the hydrogenic hybrid fluid-kinetic code SOL-KiT\cite{Mijin2021}, which has been used to tackle some of the problems detailed above, namely the quantification of kinetic electron effects on parallel transport and reaction rates in equilibrium and transient conditions. However, while SOL-KiT is a purpose-built code, ReMKiT1D has been designed as a framework such that models, like the one implemented in SOL-KiT, can be easily built, used, and modified by a variety of users. Furthermore, it allows for improvements in performance by introducing a second parallelizable dimension in the electron distribution function harmonics (see Sections \ref{MPI} and \ref{Benchmarking}).

A brief overview of problem types and equations ReMKiT1D is designed to tackle will be given in Section \ref{ProblemClasses}, after which the basic concepts required to set up ReMKiT1D simulations are discussed in Section \ref{BasicConcepts}
. Once the basic concepts are introduced, a concrete example of a simple ReMKiT1D workflow will be presented in Section \ref{PythonFortran}
, showcasing the basic idea behind coupling Python and Fortran codebases through JSON configuration files. For readers interested in the software design and implementation details behind ReMKiT1D, these are covered in Section \ref{Design}
. In order to build confidence in the software package, an extensive list of verification and benchmarking problems are presented in Section \ref{Benchmarking}, including parallel performance scaling tests with the novel parallelization in distribution function harmonics. Planned and potential use cases and extensions of the framework are discussed in Section \ref{Discussion}.

\section{Target problem classes}
\label{ProblemClasses}

With a focus on mathematical models arising in the research of transport along magnetic field lines in the SOL, ReMKiT1D has been designed to handle systems of nonlinear ODEs and PDEs arising from coupled fluid and kinetic models. These will be laid out in this section, without focusing on any individual model or implementation. 

\subsection{Systems of nonlinear ODEs}

Nonlinear ODEs arise both from the spatial and velocity discretization of fluid and kinetic models of interest, as well as naturally in the context of collisional-radiative modelling. In general, given a vector of variables $\vec{v}$, the system of interest can be written as

\begin{equation}
    \frac{d\vec{v}}{dt} = \mathbf{M}(\vec{v})\cdot\vec{v} + \vec{\Gamma},
\end{equation}
where $\mathbf{M}(\vec{v})$ is the matrix of (nonlinear) coupling coefficients and $\vec{\Gamma}$ some constant vector. It is convenient to write the system in this way both for the purposes of defining the default implicit integration scheme used, and to draw parallels with the form of general collisional-radiative models \cite{Greenland2001}. In general, the systems of ODEs that occur in problems of interest are stiff, and implicit integration schemes together with potentially flexible operator splitting might be required (see Sections \ref{TimeIntegration} and \ref{Manipulators}).

\subsection{1D PDEs}

Systems of hyperbolic conservation laws of the form

\begin{equation}
    \frac{\partial X\left(\vec{x}\right)}{\partial t} +\nabla\cdot\vec{\Gamma}_X\left(\vec{x}\right) = S_X,
    \label{eq:hyperbolic}
\end{equation}
where $X$ is a conserved quantity, $\vec{\Gamma}_X$ the flux of that quantity, and $S_X$ the source/sink of $X$, arise naturally in the modelling of multi-species fluids. Many well-known methods of solving such equations exist, reducing the system through discretization to a system of, in general, nonlinear ODEs. The complexity of the conservation laws comes from the forms of the fluxes and sources, which can be complex nonlinear functions of conserved quantities. Furthermore, cases where evolving the primitive (instead of conserved) quantities is simpler also arise. This can lead to parabolic equations, such as those that arise in heat conduction problems, as well as combined advection-diffusion problems. 

Finally, equations without an explicit time derivative term can arise, as is the case with the elliptic problem of Poisson's equation

\begin{equation}
    \label{eq:poisson}
    \Delta \varphi = f.
\end{equation}

Further complications arise with the addition of complicated boundary conditions due to the interaction of the SOL plasma with the wall, such as the sheath boundary condition and recycling. These are mentioned in \ref{appendix:SOL-SK}, and for more details the reader is encouraged to consult previous work \cite{Mijin2021,Batishchev1999,Procassini1992}, particularly in the context of kinetic boundary conditions. 

As such, it is of interest to at least attempt to cover as many of these cases as possible, which is simplified somewhat due to the 1D nature of the problems considered in this manuscript. Implementation details behind the 1D differential operators available in ReMKiT1D will be presented in Section \ref{Operators}.

\subsection{1D electron kinetic equation}

Electron kinetic effects such as heat flux suppression, sheath boundary effects, as well as potential modification of reaction rates due to non-Maxwellian distributions, are of interest both in equilibrium and transient conditions. Following the approach in SOL-KiT\cite{Mijin2021}, these effects are included as emergent physics through the option to solve the 1D electron kinetic equation, 

\begin{equation}
    \frac{\partial f}{\partial t} + v_x\frac{\partial f}{\partial x} - \frac{eE}{m_e}\frac{\partial f}{\partial v_x} = \left(\frac{\delta f}{\delta t}\right)_{c}.
\end{equation}
In principle, the electron distribution function is naturally represented in a spherical harmonic basis, which accounts well for collision symmetries with heavy particles. Assuming that the distribution function is azimuthally symmetric (around the $v_x$ direction), it can instead be expanded more simply into Legendre harmonics $f_l$, resulting in an equation of the form

\begin{equation}
    \frac{\partial f_l}{\partial t}  = A_l + E_l + C_l
    \label{eq:kinetic_l}
\end{equation}
where $A_l$, $E_l$, and $C_l$ represent spatial advection, velocity space advection due to the electric field, as well as collision operators, respectively. This enables fine control over the fidelity of the kinetic effects included, through both choosing the number of resolved distribution function harmonics, as well as controlling the individual operators evolving each harmonic. Fluid moments are obtained directly from moments of the individual harmonics, such that scalar moments are given by

$$<\phi> = \int \phi(v) f(\vec{v}) d\vec{v} = 4 \pi \int_0^\infty \phi(v) f_0(v)v^2dv,$$
and vector moments by

$$<\vec{a}> = \int \vec{a}(v) f(\vec{v}) d\vec{v} = \frac{4 \pi}{3} \int_0^\infty a(v) f_1(v)v^2dv$$
For more details on the expansion, the reader is directed towards the SOL-KiT model and other similar models in the literature\cite{Kingham2004,Bell2006,Hornsby_2010,Tzoufras2011}. Here it will only be mentioned that the capability for including the Coulomb collision operator outside of the Lorentz approximation and the proper treatment of inelastic electron-neutral collisions is desirable and has been implemented in ReMKiT1D following the techniques established in SOL-KiT\cite{Mijin2021}. 

Given the broad range of problems and the aim for flexibility, the need for separating high-level concepts and implementation arises. ReMKiT1D endeavours to exploit low-cost abstraction and the concept of scalable design to produce a framework that is both fit for purpose in terms of adequately addressing the target problem classes above, as well as providing the user with low level control, high level quality-of-life features, flexibility in the models and numerics, and rapid iteration. This requires at least a conceptual separation of high-level concepts and those directly tied to the implementation of operators and solvers in 1D, and an attempt has been made to separate these concepts with the design of the source code as well. These high-level concepts and the 1D implementation will be presented in the following two sections.

\section{Basic concepts}
\label{BasicConcepts}

In this section we cover the basic and the highest-level concepts necessary to build simulations using ReMKiT1D. We go over the four surface-level concepts used to specify simulation parameters. The section is designed to both offer a pedagogical introduction to the topic as well as act as a standalone high-level explanation of the codebase.

In order to quickly demonstrate these concepts, an example setup in Python will be presented in Section \ref{PythonFortran}, where the coupling between the Python and Modern Fortran codebases will be explained.

Software design details, as well as general implementation details will be deferred to Section \ref{Design}, and only the details necessary for understanding the four concepts and how they factor into the Python example at the end of the section will be covered in this section. 

\subsection{The four surface-level concepts}
\label{4Concepts}

We start by naively considering the building blocks of our differential equations. Firstly, we wish to solve the equations for some variables of interest. In order to do so, we need a way to represent individual terms of the equations, as well as a way to perform integration in time. Finally, our variables live on some grid, which should encode coordinate/geometrical data.

\begin{figure}[htb]
    \centering
    \includegraphics[width=0.9\textwidth]{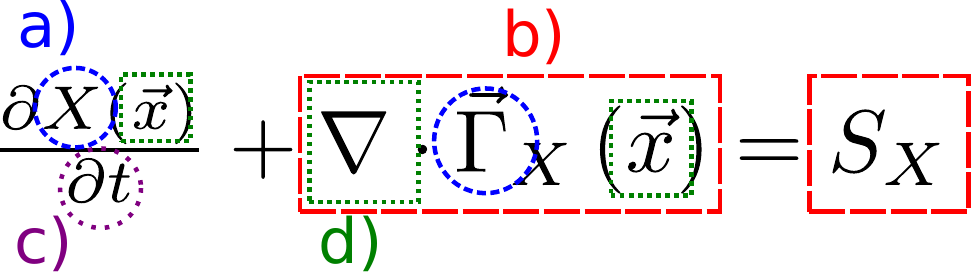}
    \caption{The four surface-level concepts highlighted on the 1D hyperbolic PDE from Section \ref{ProblemClasses}. a) dashed blue circles: The quantities we store, evolve/calculate, and output - variables. b) dashed red boxes: calculated quantities/objects used to evolve variables - terms. c) dotted purple circle: an algorithm to evolve the variables - time integration. d) dotted green boxes: information about the coordinate system - the grid.}
    \label{fig:4_concepts}
\end{figure}

Following the above reasoning, we need to represent 4 concepts:

\begin{enumerate}
    \item Variables - values of quantities we are calculating or evolving using our equations
    \item Terms - objects specifying how to evolve variables
    \item Time integration - a way to evolve our variables in time 
    \item Grid - information about the coordinates/geometry for the variables to live on and for some terms to use
\end{enumerate}

A breakdown of equation~(\ref{eq:hyperbolic}) along the above lines is shown in Figure~\ref{fig:4_concepts}. We will now look into each of these concepts in more details, examining requirements and basic implementation details so we can use them to build a simple simulation.

\subsubsection{Variables}
\label{Variables}

As noted above variables represent discretized data. Based on our target problem classes and the strive for flexibility we have the following three requirements on the variable data structure:

\begin{enumerate}
    \item Variables of different dimensionalities should be treatable. This includes scalar, fluid (1D), and distribution function (1D2V) variables.
    \item The representation of variables needs to be compatible with implicit methods so that we can safely treat stiff systems of equations.
    \item We might be interested in calculating some variables from non-integration rules. For example we might be evolving a particle flux, but need a flow speed. It should be possible to derive the flow speed from the two variables - particle flux and particle density.
\end{enumerate}

The first requirement above dictates a categorisation of variables based on their dimensionality, which is simple enough. It is the other two requirements that produce the non-trivial categorisation into implicit and derived variables.

Since implicit methods tend to involve matrix inversions/solutions of linear systems we need to be able to represent implicit variables in a way where we can use a solver library, such as PETSc\cite{balay1998petsc}. This simply means that an implicit variable needs to know how to index its values into some global vector used by the solver. If this is satisfied we can construct terms with matrix representations that can be used directly with the matrix solver (see the following subsection). 

The final requirement above translates into the concept of derivations, which is extensively used in ReMKiT1D to wrap functional dependence of some variables on others. A variable in ReMKiT1D which is not implicit can have a derivation associated with it, alongside an argument list (containing variables) for that derivation. Based on the argument list, ReMKiT1D will make sure that all MPI communication is done so that any required variables are calculated before they are used as an input to some derivation. 

An example of a derivation is the calculation tree derivation. This is essentially an expression tree, enabling the translation of Python-like expressions into objects that the Fortran codebase can use. 

In the example at the end of this section, we will add both implicit and derived variables, as well as showcase how calculation tree derivations can be used to quickly convert Python expressions into functions that are executed during the Fortran runtime.

For more details on how variables are implemented in ReMKiT1D, how they are communicated, and how tree derivations work see Section \ref{VariableContainer}.

\subsubsection{Terms and Models}
\label{Terms}

As noted at the start of this section, we need a way to represent the various terms in our equations. These equations are of the form $dn/dt=\sum_i S_i$ or $\sum_i S_i=0$, where $S_i$ are the additive terms we seek to represent. 

In ReMKiT1D, terms are always associated with a variable, which we refer to as the evolved variable. For example, in  $dn/dt=\sum_i S_i$, it is straightforward to see that $n$ would be the evolved variable. For equations of the form $\sum_i S_i=0$, we still need to associate all the terms with a single 'evolved' variable, for example $\varphi$ in equation (\ref{eq:poisson}). Variables calculated using such equations are referred to as stationary in ReMKiT1D. 

All terms in ReMKiT1D can be evaluated, i.e. their values can be calculated for a given set of input variable values, and can in turn be stored in variables. As of the writing of this manuscript (and v1.0.x of the framework), while there are many facilities implemented to treat general matrix-free terms in the future, ReMKiT1D focuses on matrix terms, since these can easily be used in the implemented implicit time integration method.

\begin{figure}[htb]
    \centering
    \includegraphics[width=0.9\textwidth]{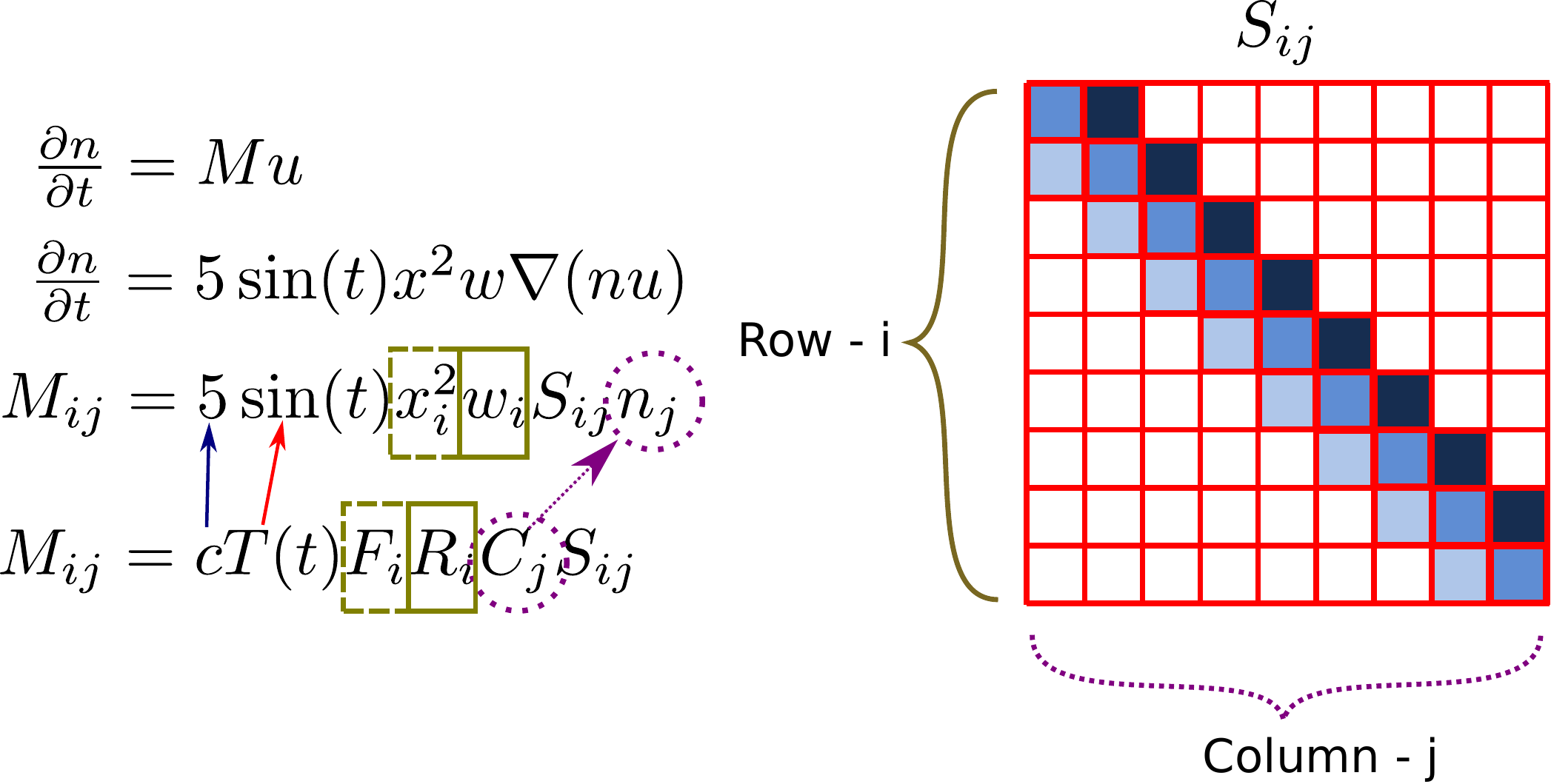}
    \caption{A toy term showcasing all of the components of an arbitrary matrix term (equation (\ref{eq:matTerm})). Left: $c=5$; $F_i=x_i^2$, where $x_i$ is the value of the $x$ coordinate associated with row index $i$; $T(t)=\sin(t)$; $R_i=w_i$, where $w$ is some variable in the ReMKiT1D context; $C_j=n_j$ because the $\nabla$ operator acts on both $n$ and $u$. Right: A sketch of a potential stencil for the $\nabla$ operator, perhaps using central differencing. Here the three different shades of blue are meant to represent different values.}
    \label{fig:matrix_term}
\end{figure}

Let both $n$ and $u$ be implicit variables, i.e. variables which have a mapping to the global solver (PETSc) vector. Then a matrix term describing the evolution of $n$ represents the RHS of 

\begin{equation}
    \frac{dn_i}{dt} = M_{ij}u_j,
\end{equation}
where summation on repeated indices is assumed. The indices $i$ and $j$ can be thought of as linearly indexing into the global solver vector and containing coordinate information (see \ref{appendix:BDE} for a formal description of the global solver vector as well as the default implicit integration scheme). The matrix $M_{ij}$, whose functional dependence on various variables, coordinates, and time is dropped here for readability, is then composed of multiplicative components and a stencil $S_{ij}$ so that

\begin{equation}
\label{eq:matTerm}
    M_{ij}=c T(t) F_i R_i C_j S_{ij},
\end{equation}
where $c$ is a constant scalar (usually associated with the normalization - see example at the end of this section). $F_i$ is a constant array that depends only on the index $i$ (and not on any variables), and thus only includes explicit multiplicative dependence on any coordinates (spatial or velocity). $T(t)$ is a time dependant scalar used to encode any explicit time dependence of the matrix term. Finally, $R_i$ and $C_j$ are vectors, which are functions of some variables in general, and which dimensionally conform to the evolved and implicit variable, respectively. In practice, these are set to products of variables raised to some power, i.e.

$$R_i = \prod_n v_{n,i}^{p_n},$$
where $v_{n,i}$ is variable $v_n$ evaluated at the coordinate set (spatial, velocity) corresponding to the index $i$. Together with stencils, these fully define a matrix term. Stencils come in different forms, representing the discretization of different operators. Some will be presented in Section \ref{Operators}, but a good example is a diagonal stencil, i.e. a Kronecker delta. 

To better illustrate the structure of a matrix term, Figure \ref{fig:matrix_term} shows a toy term and how it is mapped onto different components of $M_{ij}$. 

\begin{figure}[htb]
    \centering
    \includegraphics[width=0.9\textwidth]{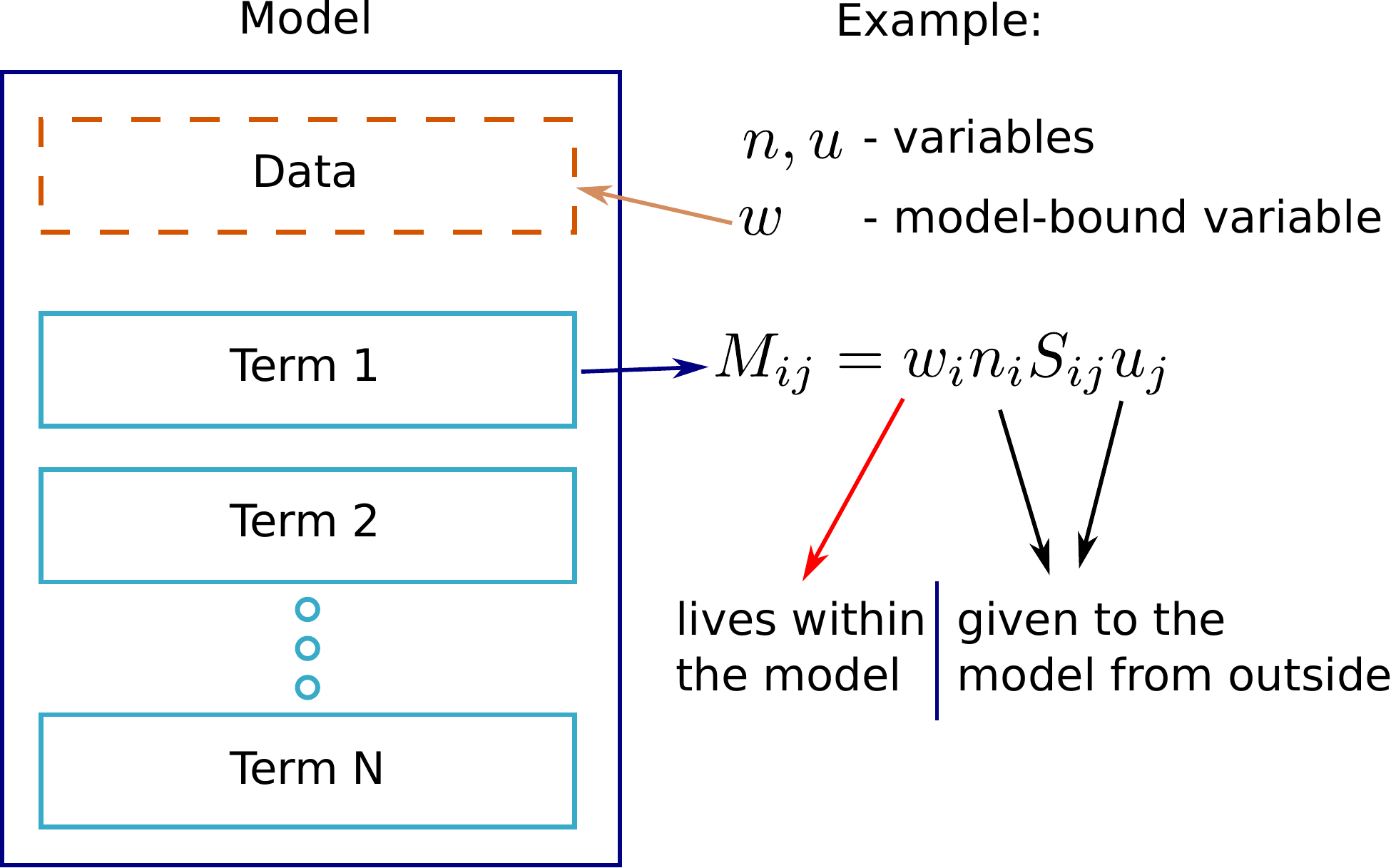}
    \caption{Left: a schematic of a model, representing groups of terms and any associated data. Right: An example contrasting variables passed to the model and data bound to the model. Here $n$ and $u$ are passed to the model in order to be used in one of its terms, while $w$ lives within the model. $w$ can then be calculated using mechanisms such as derivations. This way $w$ is encapsulated in the model object.}
    \label{fig:model}
\end{figure}

In practice, terms dealing with related physics often require the same data. An extreme example would be a collisional-radiative model, where the densities of many states might be evolved based on many different reactions and rates. Each of these rates needs to be computed, and it becomes impractical to define variables for each rate. Because of cases like these, ReMKiT1D always groups terms within model objects, allowing data to be bound to the model instead of it being passed to it in the form of variables. A schematic of a general model as well as an example contrasting model-bound data and variables passed to the model is shown in Figure \ref{fig:model}. Models are the building block of ReMKiT1D simulations, and their relationships with other components will be explored in depth in Section \ref{Design}.

\begin{figure}[htb]
    \centering
    \includegraphics[width=0.9\textwidth]{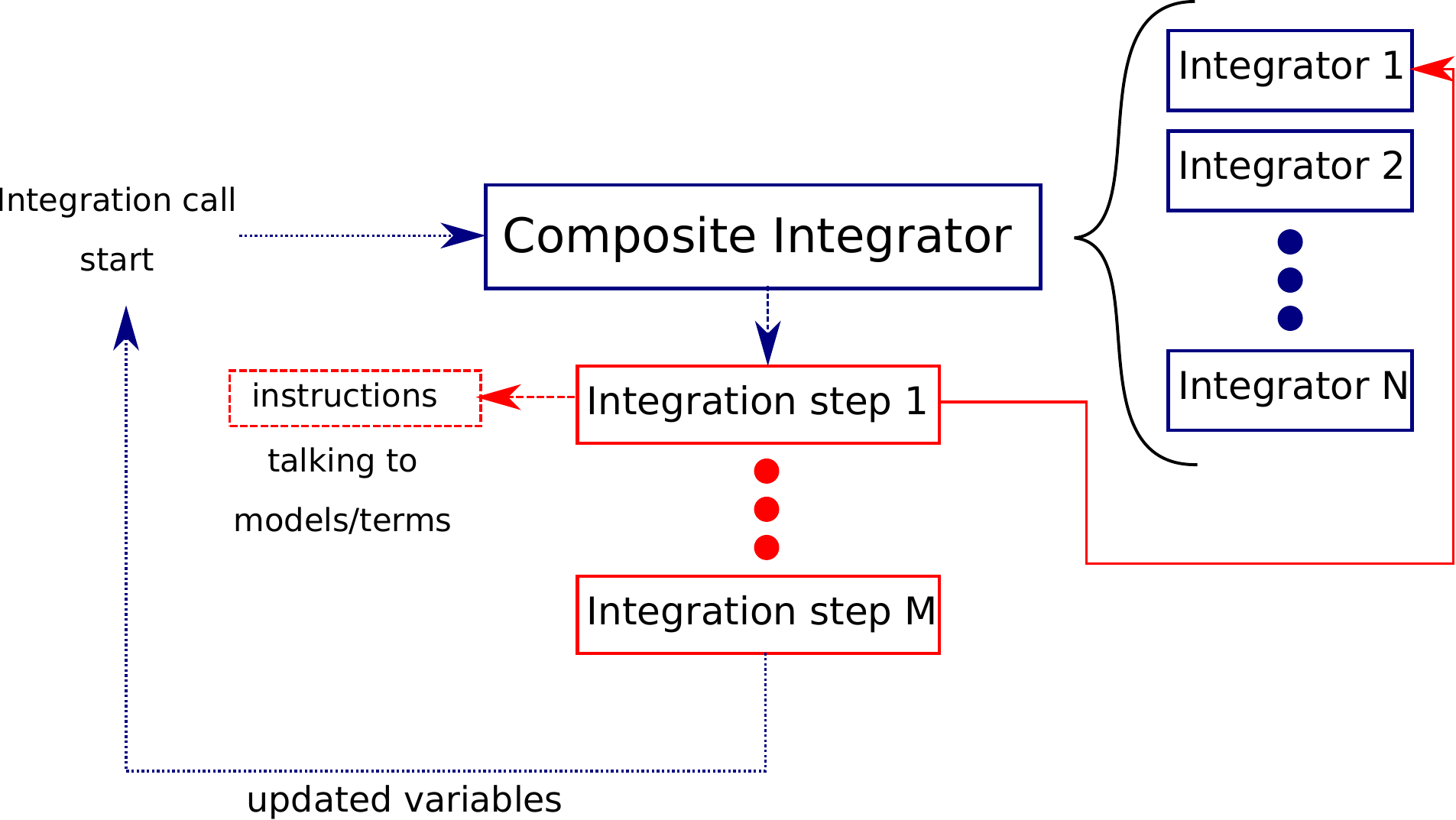}
    \caption{A simplified schematic of an integration call in ReMKiT1D. The integration call starts at the top-level composite integrator object, which steps through a series of integration steps. Each step is associated with a particular integrator (integration strategy) and instructions (e.g. relating to which models/terms the step is responsible for).}
    \label{fig:time_integration}
\end{figure}

\subsubsection{Time integration}
\label{TimeIntegration}

With the concepts of variables and models/terms introduced, the next step is to define how to treat integration in time. ReMKiT1D allows a flexible definition of integration algorithms based on built-in integrator types. Currently supported integrators are a Backwards Euler implicit integrator (see \ref{appendix:BDE}) and explicit Runge-Kutta integrators up to fourth order\footnote{Arbitrary Butcher tableau support has been implemented, but not available in the Python interface as of v1.0.x.}. 

ReMKiT1D allows the definition of integration steps, which can be associated with an integrator object, and can be instructed to integrate a subset of the terms/models. This can be useful in operator splitting methods, such as those for kinetic equations where the advection is done explicitly and collisions are treated using an implicit step \cite{Tzoufras2011}. Further control can be achieved through specifying more detailed instructions for integration steps. As such, a single integration call/time step in ReMKiT1D will step through the defined integration steps. Time-stepping is performed until some termination condition is reached (for example running a certain number of time steps). A sketch of how an integration call flows in ReMKiT1D is shown in Figure \ref{fig:time_integration}. While some of the software design details behind integrators and integration steps will be presented in the next section (see Section \ref{Manipulators}), it is useful here to introduce these concepts loosely as they are used in the example at the end of the section.

\subsubsection{The grid}
\label{Grids}

Finally, we need to represent the grid on which variables live and which is used to define self-consistent differential operators. While ReMKiT1D offers the flexibility to its users to define custom operators and grids, a convenient default is offered. This consists of a staggered spatial grid, together with a harmonic and velocity grids used when working with distribution function variables, and will be presented in this section.

\begin{figure}[h!]
    \centering
    \includegraphics[width=0.95\textwidth]{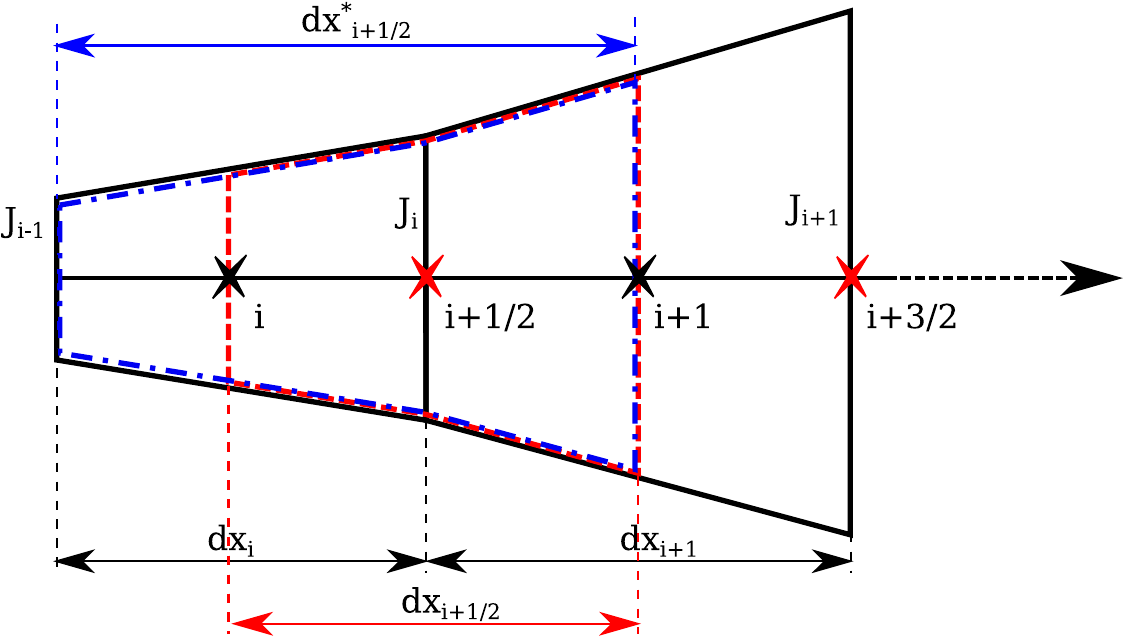}
    \caption{The default ReMKiT1D spatial grids. Black cells are cells on the regular grid, with their cell centres labelled by whole numbers $i$ and marked by black stars, and their right cell face Jacobians given by $J_i$. The dual/staggered grid points are marked with red stars and the corresponding cells with dashed red lines. Dot-dashed blue lines mark an example of an extended boundary dual cell, in this case assuming that the left face of cell $i$ in the schematic is an outer boundary face.}
    \label{fig:grid}
\end{figure}

The default spatial grid in ReMKiT1D consists of a 1D array of cells $i$. These indices can be thought of as labeling cell centres of stacked truncated cones, such that their base areas are given by cell face Jacobians $J_i$. Explicitly including variable face Jacobians allows for the representation of flux tubes of varying cross-sections in the SOL. Together with the cell widths $dx_i$ (heights of our truncated cones), these Jacobians determine cell volumes $V_i=dx_i(J_{i}+J_{i-1})/2$. Alongside the grid of cells $i$, a dual/staggered grid of cells $i+1/2$ is defined, indexing the cell edges/faces along the x-direction. The volumes and face Jacobians of these cells are calculated so that they fit into the regular grid. Thus the right face Jacobian of cell $i+1/2$ is the cell centre Jacobian/area of cell $i+1$, $J_{i+1/2}=(J_{i}+J_{i+1})/2$, and the volume is given by $V_{i+1/2}=dx_i(J_{i-1/2}+J_{i})/4+dx_{i+1}(J_{i+1/2}+J_{i})/4$. Near boundaries, dual grid cells can be extended so that the total volume of the dual grid represents the same volume as the regular grid. This is useful for applying boundary conditions to variables defined on the dual grid. A schematic of the default ReMKiT1D spatial cells is shown in Figure \ref{fig:grid}. 

Fluid variables thus live either on the regular or the dual grid, with the possibility of linear interpolation between them. As a rule of thumb, and as is common in staggered finite volume methods, scalar fields (such as the density or pressure) live in cell centres and vector fields (e.g. fluxes) live on cell edges/the dual grid. Distribution variables, on the other hand, either live entirely on the regular grid, or their even $l$ harmonics live on the regular and their odd harmonics on the dual grid, or vice-versa. This is because moments of even harmonics include densities and energies, while the odd harmonics produce moments that determine various fluxes.

When the spatial grid is periodic, the regular and dual grids have the same number of cells. Otherwise the dual grid has one fewer cell, as the outer boundaries of the domain are always assumed to be on regular cell boundaries. In terms of the actual array indexing, a variable that lives on the staggered grid with index $i$ corresponds to the right cell face of regular cell $i$, or $i+1/2$ in Figure \ref{fig:grid}.

The different harmonics in equation (\ref{eq:kinetic_l}) 
are functions of both the spatial coordinate $x$, as well as of the velocity magnitude $v$ and the harmonic index $l$. The velocity magnitude $v$ is discretized similarly to the regular spatial grid - as a 1D array of cells with specified cell centers and widths. However, unlike the spatial grid, the $v$ grid does not include explicit face Jacobians, and is instead treated more as a finite difference grid. As for the harmonic index, in general it can refer to a combination of spherical harmonic indices $l$ and $m$, even though the current implementation of ReMKiT1D uses only $l$ under the assumption of azimuthal symmetry\footnote{Support is built in for an eventual inclusion of the $m$ numbers as well}. For this reason the dimension associated with the harmonics is referred to more generally as $h$. The $h$ dimension tends to have relatively few points, reflecting the fact that usually a small number of harmonics is sufficient to resolve angular dependencies of the distribution function in many cases. Parallelization in this dimension is explained in Section \ref{VariableContainer}.

\subsection{Python-JSON-Fortran interface and example workflow}
\label{PythonFortran}

In order to demonstrate the four concepts and provide an example of how ReMKiT1D can be configured, we will cover the high-level interface of the codebase, as well as go over a simple example workflow that exercises the concepts introduced in the above sections.

ReMKiT1D's core is written in Fortran, and the code is initialized through a JSON configuration file using the json-fortran library\cite{json-fortran}. The motivation behind this approach is the combination of human readability and established IO libraries for JSON in both Fortran and Python. This section will cover IO both with JSON and HDF5 files before presenting an example of how a simple advection simulation with two equations can be generated in the Python interface. 

\subsubsection{IO with JSON and HDF5 through Python interface}

As noted above, ReMKiT1D is fully initializable using just a JSON configuration file. The JSON keys are defined in the Fortran code, and a Python interface that generates the corresponding JSON entries is provided. The main object in this Python interface is the RKWrapper, which is responsible for generating the configuration file and provides a convenient interface for creating ReMKiT1D runs without directly manipulating JSON keys. To illustrate the config.json file format, a snippet of the file produced for the advection vecrification example is provided below. The snippet contains settings for HDF5 output as well as MPI communication (more on how MPI communication is implemented in ReMKiT1D will be presented in the next section). 

\begin{lstlisting}[language=json,firstnumber=1]
"HDF5": {
        "filepath": "./RMKOutput/RMK_advection_test/",
        "outputVars": ["n", "n_dual", "T", "T_dual", "u", "u_dual", "time", "W"]
    },
"MPI": {
        "commData": {
            "haloExchangeVars": ["n", "n_dual", "u_dual", "u"],
            "scalarBroadcastRoots": [],
            "scalarVarsToBroadcast": [],
            "varsToBroadcast": []
        },
        "numProcsH": 1,
        "numProcsX": 4,
        "xHaloWidth": 1
    }
\end{lstlisting}

Data output in ReMKiT1D is performed using the HDF5 library, with HDF5 datasets generated for each variable selected for output (see above snippet for example). Each .h5 file produced this way corresponds to one time step, and the Python interface provides routines for reading these files into xarray datasets, allowing for easy data processing, including limited plotting capabilities. 

HDF5 files can also be used to initialize variable values, instead of using the initial conditions in the config.json file. Furthermore, the code can be instructed to have each processor dump a restart HDF5 file, which can then be used to restart runs from defined checkpoints. 

\subsubsection{Simple advection workflow example}

In order to illustrate the Python-level workflow involved with producing a ReMKiT1D config.json file, a simulation setup solving the following equations will be explored

\begin{align}
\frac{\partial n}{\partial t} &= - \frac{\partial u}{\partial x}, \\
    m_i \frac{\partial u}{\partial t} &= - \frac{\partial (nkT)}{\partial x},
\end{align}
where $m_i$ is here taken to be the hydrogen mass, and the flux is somewhat unconventionally written as $u$. Before moving on to the details of the workflow example, it is useful to note the default normalization scheme used in ReMKiT1D. While the users are free to set their own normalization constants for each term, the code comes with the following default normalization, used in most pre-built models, borrowing heavily from SOL-KiT\cite{Mijin2021}. 

The three independent normalization quantities are the reference density $n_0$ (in m$^{-3}$), temperature $T_0$ (in eV), and ion charge $Z_{ref}$. These are usually set to $10^{19}\text{m}^{-3}$, $10\text{eV}$, and 1, respectively. From these all derived normalization quantities are obtained:

\begin{itemize}
    \item Velocity (both for the velocity grid and flow speeds) is normalized to $v_{th}=(2eT_0/m_e)^{1/2}$
    \item Time is normalized to the reference ion-electron collision time $t_0=v_{th}^3/(\Gamma_{ei}^0n_0\ln{\Lambda_{ei}(T_0,n_0)}/Z_{ref}$, where $\Gamma_{ei}^0=Z_{ref}^2e^4/(4\pi m_e^2 \epsilon_0^2)$ and the Coulomb logarithm is taken from the NRL Plasma Formulary\cite{Huba2013}.
    \item Length is normalized to $x_0=v_{th}t_0$
    \item The distribution function units are in $n_0/v_{th}^3$
    \item The electric field is normalized to $m_ev_{th}/(et_0)$
    \item Transition energies are normalized to $T_0$
    \item Heat flux is normalized to $m_en_0v_{th}^3/2$
    \item Cross-sections are normalized to $1/(x_0n_0)$
\end{itemize}
In normalized units, the above equations become
\begin{align}
\frac{\partial n}{\partial t} &= - \frac{\partial u}{\partial x}, \\
\frac{\partial u}{\partial t} &= - \frac{m_e}{2m_i}\frac{\partial (nT)}{\partial x},
\end{align}
which is what will be implemented in the example that follows. The following code snippet initializes the wrapper and sets up IO and MPI settings:

\begin{lstlisting}[language=Python]
from RMK_support import RKWrapper, Grid, Node, treeDerivation
import RMK_support.simple_containers as sc
import RMK_support.IO_support as io

#initialize wrapper object
rk = RKWrapper()

#set IO paths
rk.jsonFilepath = "./config.json" # Default value
hdf5Filepath = "./RMKOutput/RMK_advection_test/"
rk.setHDF5Path(hdf5Filepath) # The input and output location of any HDF5 files used/generated by the code

#set MPI properties
numProcsX = 4 # Number of processes in x direction
numProcsH = 1 # Number of processes in harmonic direction 
haloWidth = 1 # Halo width in cells 
numProcs = numProcsH * numProcsX
rk.setMPIData(numProcsX,numProcsH,haloWidth)
\end{lstlisting}
The run being generated will be run on 4 MPI processes, all of which are in the spatial direction, as there are no distribution variable to be parallelized in the harmonic direction. The halo width used is the default one, as no operator requires a halo wider than one cell.

The grid is initialized using the following code
\begin{lstlisting}[language=Python]
xGridWidths = 0.025*np.ones(512) # widths of each spatial cell

#no velocity grid necessary - default values for the grids
vGrid = np.ones(1)
lMax = 0
gridObj = Grid(xGridWidths, vGrid, lMax, interpretXGridAsWidths=True)

rk.grid = gridObj
\end{lstlisting}
which initializes a grid of length $L=12.8x_0$.

Basic variables are added in the following way
\begin{lstlisting}[language=Python]
n = 1 + np.exp(-(gridObj.xGrid-np.mean(gridObj.xGrid))**2) # A Gaussian perturbation
T = np.ones(len(gridObj.xGrid)) # Constant temperature

# These will add both the variable 'v' and 'v_dual'
rk.addVarAndDual('n',n,isCommunicated=True) 
rk.addVar('T',T,isDerived=True) # isDerived removes the variable from the implicit vector
rk.addVarAndDual('u',isCommunicated=True,primaryOnDualGrid=True) # primaryOnDualGrid denotes that the main variable is u_dual, and u is interpolated 
rk.addVar('time',isDerived=True,isScalar=True)
\end{lstlisting}
After the above code, the wrapper will have the following variables registered:

\begin{itemize}
    \item 'n' - lives on the regular grid and is an implicit fluid variable (the default) - initialized as $n=1+\exp(-(x-L/2)^2)$
    \item 'n\_dual' - lives on the dual/staggered grid and is derived by linearly interpolating 'n' from cell centres onto cell edges 
    \item 'T' - a derived fluid variable with no derivation rule associated with it, effectively making it constant - initialized to 1
    \item 'u\_dual' - represents the flux, lives on the dual/staggered grid, and is an implicit fluid variable - initialized to 0
    \item 'u' - lives on the regular grid and is derived by interpolating 'u\_dual' from cell boundaries onto cell centres
    \item 'time' - an explicit scalar variable that the code will recognise as the time variable and use it in that way
\end{itemize}

To demonstrate how derivations are added, the following snippet creates a calculation tree from a Python equation (see Section \ref{TreeDerivations}
 for more details) for the normalized total energy $W$ and adds the corresponding derivation to the wrapper
\begin{lstlisting}[language=Python]
#add individual variables as nodes
nNode = Node('n')
uNode = Node('u')
TNode = Node('T')

massRatio = 1/1836 # approximate electron-proton mass ratio

#tree representation of normalized total energy calculation
wNode = 1.5*nNode*TNode + uNode**2/(nNode*massRatio) # assuming normalization to n_0*e*T_0

# Registering the derivation in the wrapper with the name "wDeriv"
rk.addCustomDerivation("wDeriv",treeDerivation(wNode)) 
\end{lstlisting}
Then one can add the variable to be calculated with this derivation with
\begin{lstlisting}[language=Python]
rk.addVar("W",isDerived=True,derivationRule=sc.derivationRule("wDeriv",['n','u','T']))
\end{lstlisting}
where the new variable 'W' is associated with the derivation rule "wDeriv" and requires the three variables that act as leaf nodes in the calculation tree. In general, the list of required variables depends on the type of derivation as well as the use case.  

At this point we can start adding the models and terms, beginning with the continuity equation and the corresponding flux divergence term
\begin{lstlisting}[language=Python]
# declare a new Model object
newModel = sc.CustomModel(modelTag="nAdvection")

#create a new general matrix term
divFluxTerm = sc.GeneralMatrixTerm(evolvedVar='n',implicitVar='u_dual',customNormConst=-1.0,stencilData=sc.staggeredDivStencil())
newModel.addTerm("divFlux",divFluxTerm) # add a new term with tag "divFlux"
rk.addModel(newModel.dict())
\end{lstlisting}
where the implicit variable is 'u\_dual' since it is the implicit flux variable, and the stencil is staggered as 'n' and 'u\_dual' live on different grids (for detailed definitions of spatial 1D stencils see Section \ref{Operators}). Similarly, the pressure gradient term is added as 
\begin{lstlisting}[language=Python]
newModel = sc.CustomModel(modelTag='pGrad')
#Required variable data for pressure 
vData = sc.VarData(reqColVars=['T']) 

gradTerm = sc.GeneralMatrixTerm(evolvedVar='u_dual',implicitVar='n',customNormConst=-massRatio/2,stencilData=sc.staggeredGradStencil(),varData=vData)
newModel.addTerm("gradTerm",gradTerm)
rk.addModel(newModel.dict())
\end{lstlisting}
where now the $C$ array in equation (\ref{eq:matTerm}) is set to the 'T' variable on line 3, making the staggered gradient stencil act on both 'n' and 'T', with the temperature value being lagged in time by the non-linear solver (has no effect in this example since it is a constant). Note that neither of the terms added has a boundary condition term. Since the grid is staggered, the main boundary conditions are on the divergence of the flux, and with no boundary condition term specified, these default to reflective boundary conditions (0 flux on boundaries).
The only remaining setup concerns the time integration, starting with the definition of the implicit integrator and its addition to the composite integrator object
\begin{lstlisting}[language=Python]
# the implicit BDE integrator that checks convergence based on the variables 'n' and 'u_dual'
integrator = sc.picardBDEIntegrator(nonlinTol=1e-12,absTol=10.0,convergenceVars=['n','u_dual']) 

rk.addIntegrator("BE",integrator)
\end{lstlisting}
where the absolute solver tolerance is in units of machine precision\footnote{More precisely, in the units of the Fortran intrinsic function epsilon() applied to the default ReMKiT1D real variable kind.}, and the global integrator properties
\begin{lstlisting}[language=Python]
# fixed timestep in this example
initialTimestep=0.1 # in normalized time units
rk.setIntegratorGlobalData(1, # number of allowed implicit term groups - grouping everything into one group per model
                           1, # number of allowed general term groups
                           initialTimestep) 
\end{lstlisting}
where no time step control is specified (see \ref{appendix:BDE} for example of time step length control), making the time step constant - $\Delta t = 0.1$. All models are also set to allow only a single term group, as there are no diagnostic variables that would require evaluating individual terms or term groups, and there is no operator splitting in the integration of individual models. In this simple example, a single integration step in the composite integrator is used, set in the following way
\begin{lstlisting}[language=Python]
# a single integration step evolving all models
bdeStep = sc.IntegrationStep("BE")

for tag in rk.modelTags():
    bdeStep.addModel(tag)

rk.addIntegrationStep("StepBDE",bdeStep.dict())
\end{lstlisting}
Finally the number of time steps is set\footnote{It is possible to set different time-stepping modes and output modes such as running until a certain elapsed normalized time is reached}, as well as how often the code should output variable data
\begin{lstlisting}[language=Python]
rk.setFixedNumTimesteps(10000)
rk.setFixedStepOutput(200)
\end{lstlisting}
The configuration file is then written by simply calling
\begin{lstlisting}[language=Python]
rk.writeConfigFile()
\end{lstlisting}
The configuration file is then ready for use. Once the output files are generated they can be loaded into an xarray Dataset object using a provided Python routine. The output of this advection test will be analyzed in Section \ref{Benchmarking}, and the Jupyter notebook with the test and analysis is available in the ReMKiT1D-Python repository.

\section{Software design and implementation}
\label{Design}

In this section we go deeper into the software design aspects behind ReMKiT1D, as well as into the concrete implementation details behind features such as calculation trees, MPI communication, as well as some specific stencils.

We begin with a descriptive presentation of the high-level concepts behind the design of ReMKiT1D. Some of these concepts overlap with the four surface-level concepts introduced in the previous section. To make it clear that we are now talking about classes and objects in an object-oriented design, both classes and objects are written in \textbf{PascalCaseBold}. 

Where appropriate, Section \ref{PythonFortran} and the example workflow there will be referenced to provide additional concrete examples of concepts discussed in this section.

\subsection{The Modeller-Model-Manipulator pattern}
\begin{figure}[htb]
    \centering
    \includegraphics[width=0.9\textwidth]{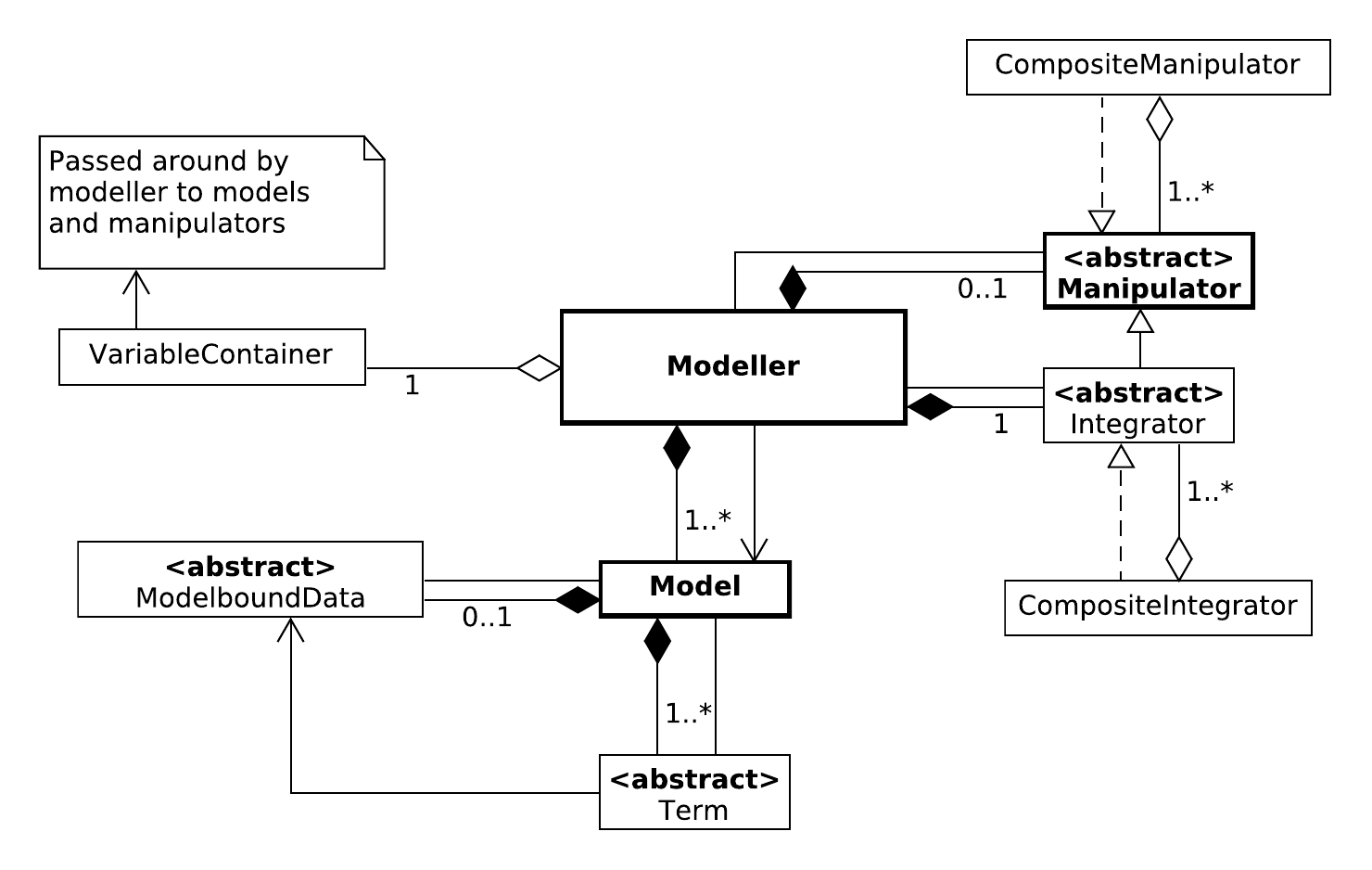}
    \caption{A simplified UML diagram of the high-level structure of ReMKiT1D, showcasing the \textbf{Modeller}-\textbf{Model}-\textbf{Manipulator} pattern, with those three components in bold edged boxes. A \textbf{Modeller} has one or more \textbf{Models} and is coupled with its \textbf{Manipulator} component so that the \textbf{Manipulator} can call \textbf{Modeller} functions and modify the variables. A special case of the \textbf{Manipulator} is the \textbf{Integrator}, which uses the \textbf{Modeller}-\textbf{Model} interface to evolve variables in time.}
    \label{fig:RMK_uml}
\end{figure}

ReMKiT1D's high-level algorithm is built on the extension of the puppeteer pattern as presented by Rouson et al \cite{rouson2011scientific}. In the original puppeteer pattern, it is the puppeteer's responsibility to manage the interaction between models on a global level. In our pattern, this responsibility is, in principle, extended to a third object, to which the additional responsibility of performing non-trivial transformations of various variables is also delegated. This third object follows the strategy pattern\cite{rouson2011scientific}
, enabling further flexibility through what is effectively dependency injection. Figure \ref{fig:RMK_uml} shows a simplified UML diagram of this pattern, as implemented in ReMKiT1D.

The fundamental object in the pattern is a central \textbf{Modeller} object, containing the variables that should be accessible to multiple components, as well as any supporting library wrapper routines (such as MPI or PETSc).

\begin{figure}[!t]
    \centering
    \includegraphics[width=\textwidth]{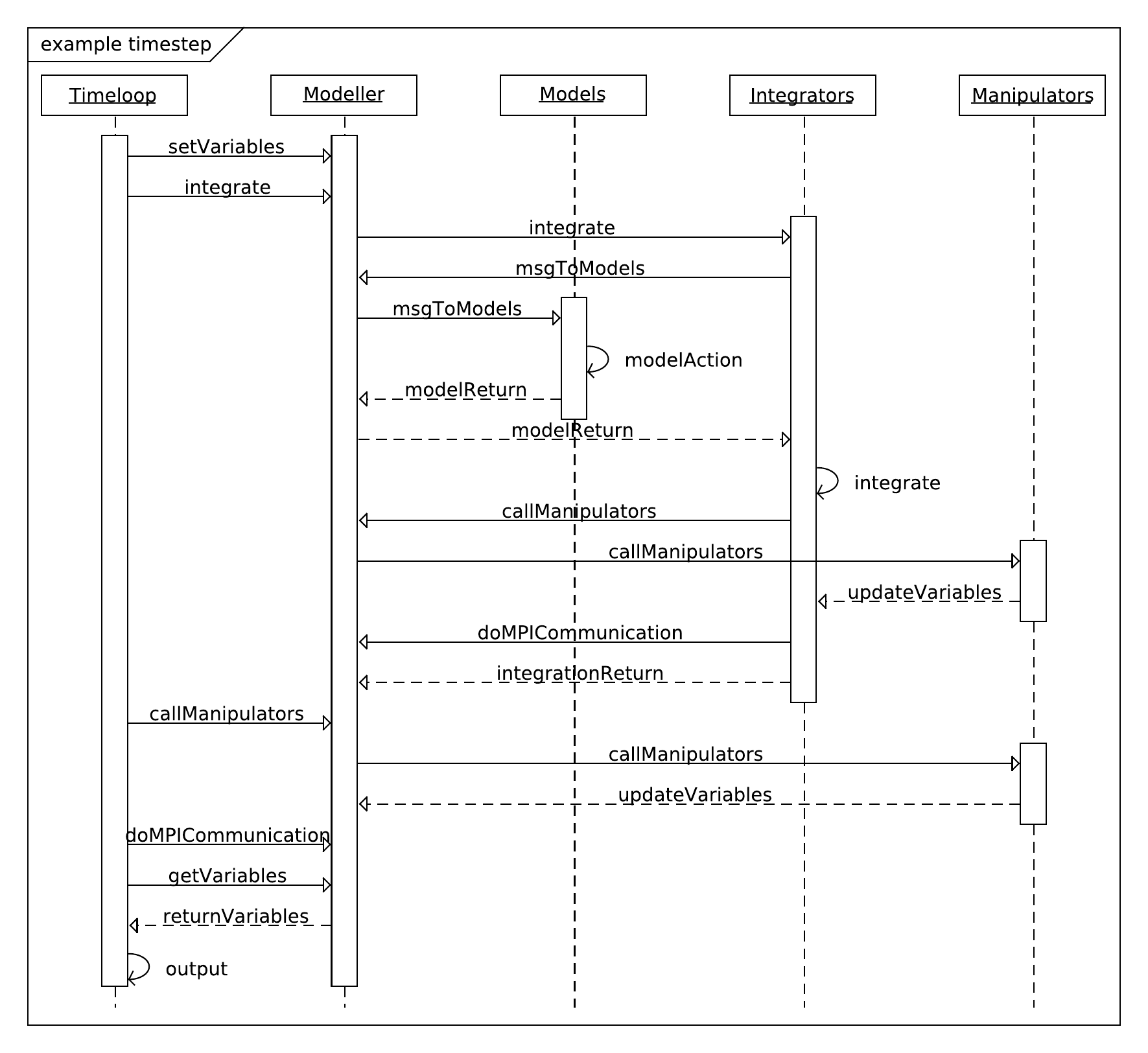}
    \caption{A simplified UML sequence showing an example time step and the associated communication between the main components of the Modeller-Model-Manipulator pattern. The external loop over time is represented here as a \textbf{Timeloop} object sending messages to the \textbf{Modeller}. Here the time step begins with setting some initial values in the \textbf{Modeller} before requesting an integration step. The integration request is forwarded to the \textbf{Integrators} which request some actions from the \textbf{Models} via the \textbf{Modeller}. For example, these might be evaluations of \textbf{Term} objects. The integration scheme might also require calling other \textbf{Manipulators} via the \textbf{Modeller} and will likely require MPI communication call requests from the \textbf{Modeller}. \textbf{Disclaimer:} This UML sequence is not an exact representation of the code, and is used primarily for illustrating the use of the high-level pattern. The names of calls/messages do not necessarily correspond to those in the code.}
    \label{fig:RMK_uml_seq}
\end{figure}

The relationships between the \textbf{Modeller} and the \textbf{Models} and \textbf{Manipulators} will be discussed in more detail in the next two subsections, but the following short summary should illustrate the responsibilities of the different pattern members:
\
 \begin{itemize}
    \item The \textbf{Modeller} provides an interface for central execution of calls such as integration (in time) and communication. The variables live in the \textbf{Modeller} and it provides access to them both to its own components as well as its users. This way, the \textbf{Modeller} can be fitted into an external loop over time, as shown in Figure \ref{fig:RMK_uml_seq}. It contains the \textbf{Models} and \textbf{Manipulators}, the interactions of which it is responsible for coordinating.
    \item The \textbf{Manipulator} modifies the variables in the \textbf{Modeller} (or local copies of those variables - see Figure \ref{fig:RMK_uml_seq}
     and description in the text below) based on calling various \textbf{Modeller} routines. This can be anything from performing an integration step while obeying communication rules to evaluating and storing individual terms into diagnostic variables. The main concept behind the \textbf{Manipulator} class is the enabling of high-level dependency injection.
    \item As presented in Section \ref{Terms}, \textbf{Models} can be thought of as collections of \textbf{Term} objects as well as potentially some data (\textbf{ModelboundData}) accessible by default only to those \textbf{Terms}. We recap the general definition of \textbf{Terms} here for convenience:  At a high level, \textbf{Terms} represent additive terms in the various equations of Section \ref{ProblemClasses}. For example, the divergence of the flux $\vec{\Gamma}_X$ in equation (\ref{eq:hyperbolic}) is an additive term acting to evolve the variable $X$. In this way, each term is associated with an evolved variable, and these variables are the same ones that live in the \textbf{Modeller}. Note that each individual \textbf{Term} only provides part of the contribution going towards the evolution of its associated variable, and it is only the \textbf{Manipulators} (and in particular \textbf{Integrators}) that can use this information to update variables in the \textbf{Modeller}.
\end{itemize}

Figure 7 \ref{fig:RMK_uml_seq} shows an example call sequence representing a single time step utilizing the pattern presented here. The steps in it can be summarized as:

\begin{enumerate}
    \item The external object (the \textbf{Timeloop} in this case) sets some initial values for the variables if needed
    \item The \textbf{Timeloop} requests an integration step from the \textbf{Modeller}, and this request is forwarded to the \textbf{Integrator} component of the \textbf{Modeller}
    \item Based on its configuration, the \textbf{Integrator} requests some actions or data from the \textbf{Models} via the \textbf{Modeller}.
    \item The \textbf{Models} perform any necessary action (this might be anything from updating their \textbf{ModelboundData} to evaluating the \textbf{Terms} contained in the \textbf{Models})
    \item Once control is returned to the \textbf{Integrator} it might request actions from the \textbf{Modeller} such as MPI communication or \textbf{Manipulator} calls\footnote{The \textbf{Manipulator} calls from within \textbf{Integrator} objects usually modify a copy of the variables from the \textbf{Modeller} local to the \textbf{Integrator}.}. 
    \item When \textbf{Manipulator} calls are requested, variables passed to it are updated based on the nature of the \textbf{Manipulator}. This can involve further calls to \textbf{Modeller} routines from within the \textbf{Manipulator}
    \item Given that \textbf{Integrators} are special cases of \textbf{Manipulators}, at the end of the integration call the variable values in the \textbf{Modeller} are updated and can be requested by the user of the \textbf{Modeller}. In this case it would be the \textbf{Timeloop}
\end{enumerate}
The above steps are not necessarily representative of the actual control flow in ReMKiT1D, and do not include the setup of the \textbf{Modeller} or other components. Instead, they are meant to illustrate how the Modeller-Model-Manipulator pattern can be used. The following subsections will now go further into detail of the components of this pattern, with the exception of \textbf{Models} and \textbf{Terms}, which have been introduced in sufficient detail in the previous section.

\subsubsection{Manipulators and Integrators}
\label{Manipulators}

The fundamental idea behind \textbf{Manipulators} is formalizing high-level dependency injection through enabling callbacks to the \textbf{Modeller}\footnote{As mentioned above, one can also think of this as an implementation of the well-known strategy pattern.}. \textbf{Manipulators} then allow for the direct manipulation of variable data in the \textbf{Modeller} (or local copies of that data in other objects). This is then the only way variables are allowed to change from within the \textbf{Modeller}. They can, however, still be set from outside, which is used in initialization.

Manipulators are stored in a \textbf{CompositeManipulator} object, which the \textbf{Modeller} calls directly, and each \textbf{Manipulator} is also associated with a priority (0 being the highest). This way, one can control when certain \textbf{Manipulators} are called. For example, one might want to call a \textbf{Manipulator} as often as possible as it modifies a variable that is used in some internal iteration of an integrator. On the other end of the spectrum, one might want to just call the \textbf{Manipulator} before outputting data, if the \textbf{Manipulator}'s task is extracting diagnostic variables such as term evaluation.

Several important data access \textbf{Manipulators} are implemented at the moment. They include term evaluation \textbf{Manipulators}, that store the evaluation value of a \textbf{Term} in one variable, useful for analysis, as well as debugging. Similarly, an extraction \textbf{Manipulator} is available for accessing \textbf{ModelboundData} values that can fit into regular variables.

Finally, the most important \textbf{Manipulators} are the \textbf{Integrators}, which have their own specialized container in the \textbf{CompositeIntegrator}. The essentials of \textbf{Integrators} were covered in Section~\ref{TimeIntegration}, where the concepts of integrators and integration steps are introduced. In the following few paragraphs some of these concepts will be discussed in slightly more detail. The \textbf{CompositeIntegrator} controls any single integration call in two ways:

\begin{enumerate}
    \item By applying any global time step control 
    \item By calling individual \textbf{Integrator} components in accordance with precisely defined integration steps
\end{enumerate}
The application of the global time step control is done by re-scaling the initial time step in accordance with some rule (see \ref{appendix:BDE} for an example). 

Integration steps are defined by the following:

\begin{itemize}
    \item The associated \textbf{Integrator} object
    \item The fraction of the global time step (the total time step requested in the integration call) associated with the step
    \item Evolution and update rules (e.g. which \textbf{Models} and \textbf{Term} groups should be evaluated or how often to update non-linear terms)
\end{itemize}
We note here that in ReMKiT1D \textbf{Terms} within \textbf{Models} can be further grouped into subgroups, which is what "\textbf{Term} groups" refers to in the above text. In combination with the grouping of \textbf{Terms}, integration steps give the user full control over any potential operator splitting through simply defining each step in sequence, and associating the \textbf{Models} (and optionally even individual \textbf{Term} groups) to be used in the evolution. Furthermore, the control over update frequency of both individual non-linear \textbf{Terms} as well as \textbf{ModelboundData} opens up performance optimization opportunities at the expense of accuracy, all accessible at the highest level of the interface. However, sensible defaults are supplied so users interested only in the high-level use of the framework do not get bogged down in details. For example, in the Python workflow presented above no \textbf{Term} sub-grouping is done, and the default behaviour of updating non-linear term as often as necessary is adopted, simplifying the setup of the integration.

\subsection{Variable containers, Derivations, and communication}
\label{VariableContainer}

Variables are stored in \textbf{VariableContainer} objects, with the main one living within the \textbf{Modeller}. The actual data structure used to store variables is an array of arrays, with multidimensional variables (distributions) flattened. A \textbf{VariableContainer} is equipped with routines to generate local (in the MPI sense) flattened vectors of all implicit variables for use in PETSc routines. 

As noted in Section 3.1.1 \ref{Variables}, ReMKiT1D supports the definition of function wrappers, referred to as \textbf{Derivations}, which take in a list of variables as arguments and produce new variable values\footnote{The implementation of the interface is a little different, with \textbf{Derivations} taking in data and indices associated with the variables that are required in the array of arrays data structure stored in the \textbf{VariableContainer}}. A derivation rule is then a combination of a \textbf{Derivation} and a list of required variables.

In general, \textbf{Derivations} wrap impure functions, allowing for changes to the internal state of the \textbf{Derivation} object. However, most derivations available in ReMKiT1D are written avoiding side-effects, with some tree-based calculation derivations, covered below, written explicitly with pure functions to enable compiler optimization. 

While the list of available \textbf{Derivation} classes is too long to cover here, it is worth noting that they can be combined both additively and multiplicatively through corresponding composite \textbf{Derivation} objects. More involved examples include derivations that take moments of distribution function variables, or specialized derivations for polynomial functions of multiple variables (see \ref{appendix:SOL-SK} for an example of where this is useful). 

\subsubsection{MPI communication and communication-safe derivations}
\label{MPI}

\begin{figure}[h!]
    \centering
    \includegraphics[width=0.95\textwidth]{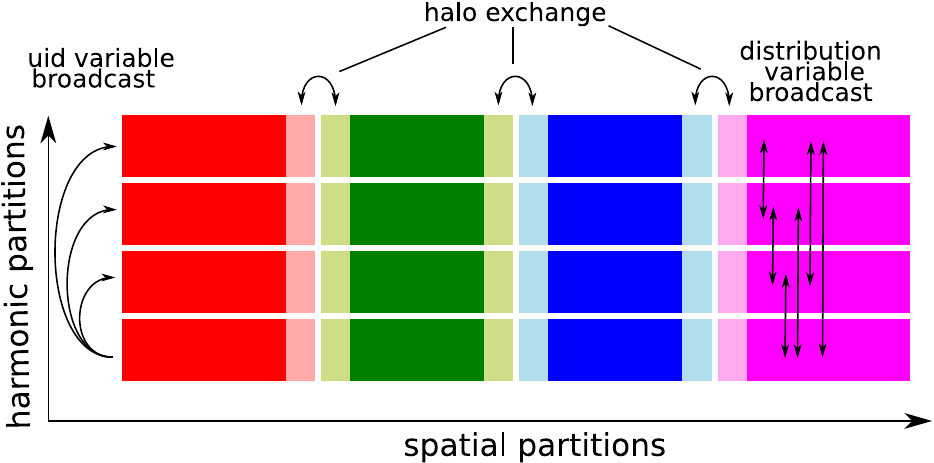}
    \caption{Schematic of a spatial-harmonic domain decomposition. Each cell represents one MPI process, and each column (shown in different colours) corresponds to a single set of spatial coordinates. Within a column fluid variables live in the first process, while harmonics are distributed across the column processes. Three of the four types of communication in ReMKiT1D are shown: halo exchange - exchanging spatial information along rows; fluid variable broadcasts - broadcasting from each column's root process to other processes in the column; distribution variable broadcast - broadcasting data from all column processes to all other processes in the same column.}
    \label{fig:mpi}
\end{figure}

ReMKiT1D utilizes MPI parallelism, with each rank responsible for evolving/calculating its own local variable data in the following way. The spatial domain is simply decomposed, with halo exchange coupling the different spatial partitions. However, in order to obtain speedup when multiple distribution harmonics are included, ReMKiT1D also allows partitioning in the harmonic domain, though in a less efficient manner due to some fundamental constraints. Thus, the domain decomposition when evolving distribution variables is inherently 2D. Each column represents a spatial partition and each row represents a harmonic partition. Hence, spatial(harmonic) information is exchanged between columns(rows). This is shown in Figure \ref{fig:mpi}.

The following rules for local variable data are observed:

\begin{itemize}
    \item Fluid variables live in the first row of each column, where they are evolved/calculated and broadcast to other processes in their respective column. However, derived variables need only be communicated if necessary (as deemed by the user and set by the problem), and are calculated on each process independently following the communication-safe algorithm as presented at the end of this subsection.
    \item Distribution variables are spread across all processes, with their harmonics partitioned within processor columns. Due to how some operators might require the entirety of the distribution function at a single spatial location, harmonics within a column are broadcast to all processors in that column.
    \item Scalar variables are assigned a host processes, such that they are broadcast from that process to all others. An example is extracting the value of a variable in the last spatial cell and passing it to all other processors, which is useful in practice when some boundary condition needs to be known by the entire spatial domain. 
\end{itemize}

Based on the above rules for variables, four types of variable communication exist in ReMKiT1D, with three shown in Figure \ref{fig:mpi}. These are

\begin{itemize}
    \item Halo exchange - where both fluid and distribution variables are exchanged within their respective processor rows
    \item Fluid variable broadcast - where fluid variables are broadcast from each processor column's root to the other column processes
    \item Distribution variable broadcast - where distribution variable harmonics are broadcast from the processes that evolves them to all other processes in the same column
    \item Scalar variable broadcast - broadcast from single host process to all others (not shown in Figure \ref{fig:mpi})
\end{itemize}

Other minor communication routines exist, primarily to check integrator convergence criteria by performing a logical reduction operation over all processes.   

The two domain decomposition dimensions are not equivalent, as might be expected from a standard 2D domain decomposition. In particular, harmonic decomposition is more communication-heavy. However, due to the large number of practically dense matrix operators, such as collision operators, speedup can be gained even with this increased cost of communication, as will be shown in Section \ref{Benchmarking}.

\begin{figure}[htb]
    \centering
    \includegraphics[width=0.9\textwidth]{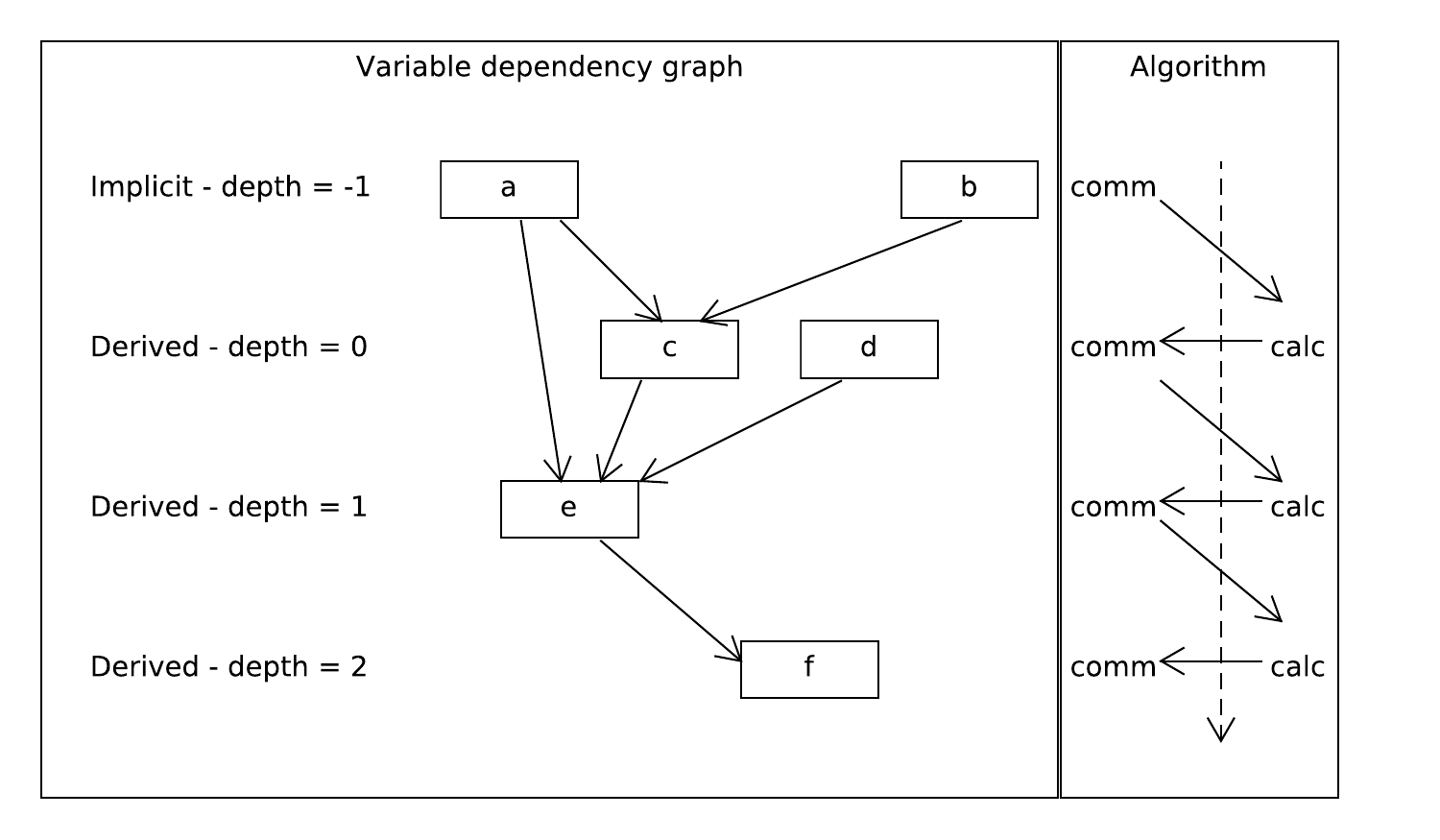}
    \caption{An example of how ReMKiT1D handles the calculation and communication of derived variables. \textbf{Left}: A graph representation of variable dependencies in this example. Here a and b are implicit variables and can be safely communicated first. c is derived using only implicit variables a and b and is thus given depth 0; d is not an implicit variable, but does not have a derivation rule associated with it (perhaps being evolved explicitly or filled in by a \textbf{Manipulator} call), and is also given depth 0. e and f depend on depth 0 and 1 derived variables, and thus have depths 1 and 2, respectively. \textbf{Right}: The derivation-safe communication algorithm - derivations are called on each depth only once the depth before it has been communicated, ensuring no out-of-order operations. \textbf{Example:} If c requires taking the central difference derivative of a, the halo of variable a must first be communicated.}
    \label{fig:derivations}
\end{figure}

It is worth noting here that some \textbf{Derivations} might require knowledge of data calculated or evolved by MPI processes other than the local one. An example would be a central difference operator derivation, which will require knowledge of spatial halo values before being able to correctly calculate the difference. Thus, there is danger of out-of-order communication and derivations. This is true in particular with scalar variables, which are always associated with a primary host process, and are broadcast to all others. If a \textbf{Derivation} on one process requires a scalar variable living on another process, the broadcast \textbf{MUST} happen before the derivation call in order for the correct value to be used. In ReMKiT1D, this is ensured by associating a derivation depth with every variable, and always using a communication-safe derivation call, as follows:

\begin{itemize}
    \item Implicit variables are given a derivation depth of -1; they are always safe to communicate and are the first to be communicated.
    \item Derived variables that only require implicit variables (or don't have any required variables) are given a depth of 0; they can be calculated once implicit variables have been communicated.
    \item All other derived variables are given depth equal to $d+1$, where $d$ is the highest depth among variables required by the derivation rule of the derived variable in question. Thus, variables of depth $d$ are calculated only after variables of depth $d-1$ have been communicated.
\end{itemize}
The above algorithm ensures safe calls to all \textbf{Derivation} routines, under the condition that there are no cyclical dependencies. This can be represented by a directed acyclic graph, as shown in Figure \ref{fig:derivations}. It is the responsibility of the \textbf{Modeller} to centralize both the communication and derivation calls through the application of this algorithm.

Finally, it is worth noting that a variable-like \textbf{ModelboundData} object is available, which stores derived variables only. Those variables, unlike in the \textbf{VariableContainer} can also represent single harmonics (variables depending only on the $x$ and $v$ coordinates), making them useful for some kinetic algorithms (see \ref{appendix:SOL-SK}). However, the above communication-safe derivation call is not available, so care should be taken that any derived variables in such a \textbf{ModelboundData} object are added in the correct order. In practice this is rarely an issue, since most derived variables in \textbf{ModelboundData} tend to be calculated using variables in the \textbf{VariableContainer}, while only a minority require variables that only live in the \textbf{ModelboundData}.

\subsubsection{Tree-based calculation Derivations}
\label{TreeDerivations}

As was presented in the example workflow in the previous section, ReMKiT1D allows for the translation of some Python expressions into \textbf{Derivation} objects directly. This is done by leveraging a tree representation of the Python expression. The expression tree is composed of nodes, where each node (represented in the Fortran code by the \textbf{CalculationNode} class) can have a particular set of properties:

\begin{itemize}
    \item Whether the node is additive or multiplicative with respect to the results of its children
    \item A single constant to add to/multiply the results of the children, depending on whether the node is additive or multiplicative
    \item An associated variable name from the \textbf{VariableContainer} - only relevant for leaf nodes, where it is treated as the result of the node's non-existent children
    \item A unary transformation, to be applied to the result of the node
\end{itemize}
\begin{figure}[h!]
    \centering
    \includegraphics[width=0.95\textwidth]{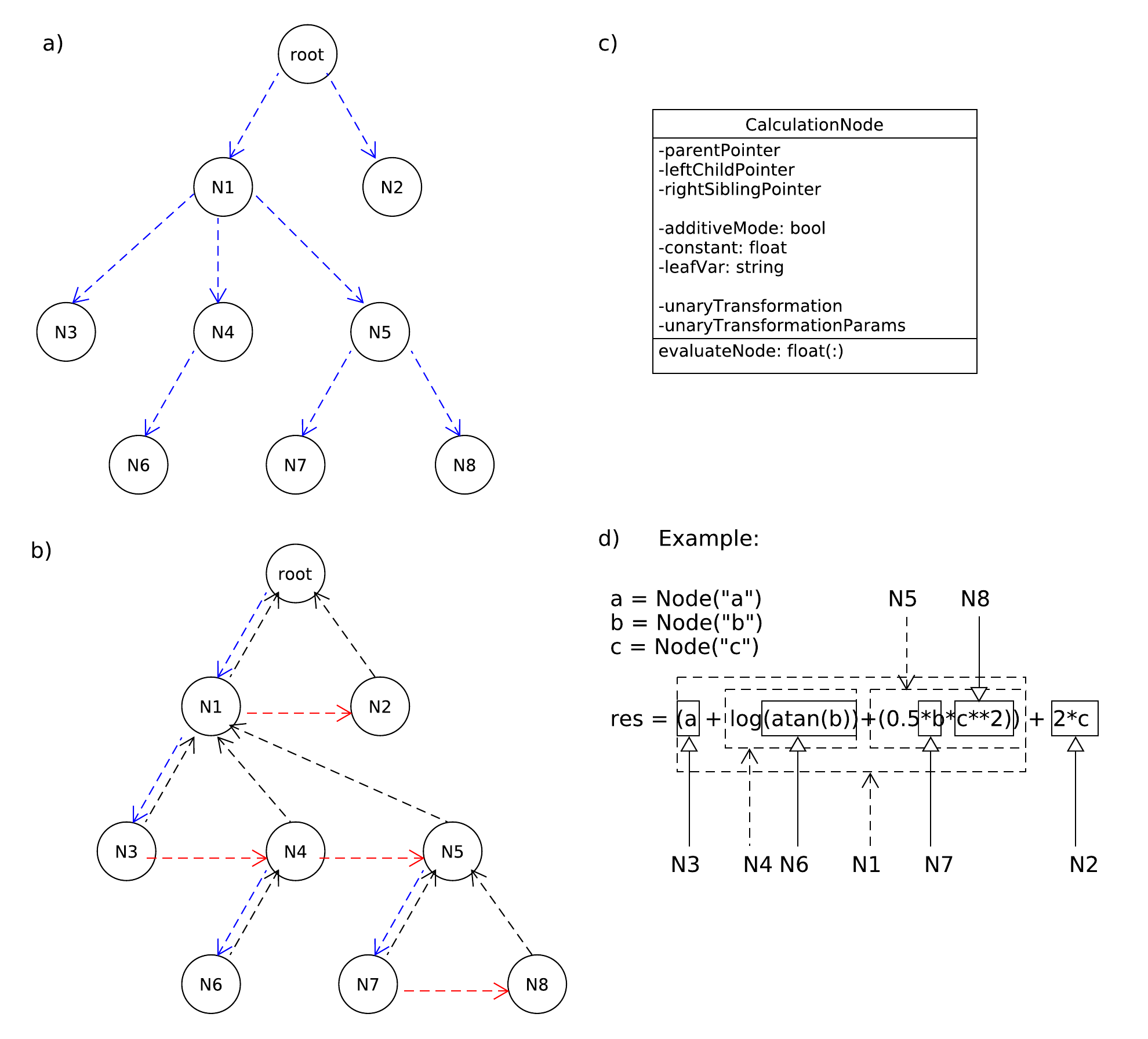}
    \caption{Example of a calculation tree. a) General expression tree diagram we are looking to represent. Each node should add/multiply the results of its children and potentially apply some unary transformation to that result. b) Left-child right-sibling representation of the calculation tree. Blue arrows connect the parents with their leftmost child, red connect nodes with their sibling to the right, and black arrows point back to the parent. c) UML representation of the \textbf{CalculationNode} class. See text for more details. d) Example of a Python expression that would generate the example tree for ReMKiT1D to use. Leaf nodes are pointed out using solid line arrows and boxes, and the composite nodes with dashed line arrows/boxes. The root node and nodes N1,N4, and N5 are two additive nodes, a node with an applied function, and a multiplicative node with a constant, respectively. Note that, in this case, all three variables participating in leaf nodes must conform, i.e. be of the same size (e.g. all must be fluid variables or distribution variables, but not a mix).}
    \label{fig:calculation_tree}
\end{figure}
By default, nodes are multiplicative, with a constant of unity, and no unary transformation. Most basic functions, such as $exp$ and $log$, are implemented as unary transformations. Unary transformations can also have associated parameters, allowing for added flexibility. An example is raising variables to integer- or real-valued powers, which are transformations where the power is a transformation parameter. Another useful parameterized transform is the \textit{shift} transform, which cyclically shifts the flattened array representation of a variable/node evaluation result some number of entries. This allows for representation of finite difference/volume operators through appropriate combinations of \textit{shift} transforms. Transforms are also supplied that can be used for the contraction of distribution variables into variables that are only defined on the spatial grid, and vice-versa. In this way, preparation work has been done for general non-matrix \textbf{Terms} to be implemented with a simple Python level interface in the future.

The result of a node evaluation is given by

\begin{equation}
    R_i = f_i(c_i+\sum_{j\in C_i} R_j),
\end{equation}
for an additive node and 
\begin{equation}
    R_i = f_i(c_i\prod_{j\in C_i} R_j),
\end{equation}
for a multiplicative node, where $f_i$ is the unary transformation, $c_i$ is the constant, and $C_i$ is the list of the children of node $i$.
The expression tree is represented in the Fortran code using a left-child right-sibling tree structure, where each child node also has a reference to its parent. The two above expressions then become

\begin{equation}
    R_i = f_i(c_i\cdot_iR_{L_i}\cdot_{P_i}R_{S_i}),
\end{equation}
with $\cdot_i$ now representing the binary operation (addition or multiplication) associated with node $i$, $L_i$ the left child of $i$, $S_i$ the right sibling of $i$, and $P_i$ the parent node of $i$. Both representations of the tree structure, as well as the UML element for the \textbf{CalculationNode} class, and an example Python expression generating an expression tree are given in Figure \ref{fig:calculation_tree}. 

A few notes on the implementation details of this particular type of \textbf{Derivation} are in order, as there are a number of Fortran peculiarities that need to be addressed. Firstly, copying derived types with pointer components is error prone, as the copy's pointers will not point to the same object as the original's pointers, but to the pointers themselves. This can lead to surprises when objects go out of scope, causing all copies of them to have their pointer components dangling. This can be avoided using Fortran's allocatable components. However, since the unary transformations are procedure pointer components, it is impossible to completely avoid this issue of dangling pointers. The way ReMKiT1D's implementation gets around these behaviours is through flattening the expression tree, and unpacking it only once evaluations are required. One can argue that the procedure pointer issue can be avoided by checking association at each \textit{evaluateNode} call, and associating the pointer with the correct procedure if it became un-associated. Unfortunately, this is technically a side-effect, and would disallow the use of pure \textit{evaluateNode} functions, which was one of the aims when designing this particular expression tree representation.

\subsection{Operators in 1D}
\label{Operators}

In this section, a number of operators will be reviewed in order to provide further implementation details, with the aim to provide concrete examples of stencils used in the codebase. However, this will not be an exhaustive list, in particular when it comes to various collision operator stencils and kinetic boundary conditions, the implementation of which is adopted from SOL-KiT\cite{Mijin2021}. 

The simplest stencil is the diagonal stencil $S_{ij}=\delta_{ij}$. It does, however, allow for the specification of evolved spatial cells/harmonics/velocity cells, effectively including only the relevant diagonal terms in the matrix. The diagonal stencil also automatically determines whether interpolation/extrapolation is necessary. This is important when staggered grids are used and the evolved and implicit variable are not defined on the same grid. In this case the diagonal stencil will automatically linearly interpolate from the implicit variable's grid onto the evolved variable's grid.

The second most important stencil group are the various spatial difference stencils, starting with the central difference stencils. Here both the implicit and the evolved variable must be defined on the same grid (regular or dual), and the implicit variable is interpolated onto the corresponding cell boundaries. An example is the central difference divergence stencil, where $i$ and $j$ here are spatial cell indices

\begin{equation}
S_{ij}u_j = \frac{J_{i}u_{i+1/2}-J_{i-1}u_{i-1/2}}{V_{i}},
\end{equation}
where $u_{i+1/2}$ is the variable $u$ interpolated on the right boundary of cell $i$. For variables defined on the dual/staggered grid, the interpolation is performed accordingly, and the above expression can be used with $i \rightarrow i+1/2$. For a gradient operator, the Jacobians are ignored, and the stencil becomes simply

\begin{equation}
S_{ij}u_j = \frac{u_{i+1/2}-u_{i-1/2}}{dx_i}.
\end{equation}
In the case of differences on a staggered grid, where the implicit and evolved variable live on different grids, the expressions look the same, but now $u_{i+1/2}$ is not interpolated, since it is the actual value of the implicit variable, considering it lives on the boundaries of cell $i$. In practice, this amounts to forward/backward difference, depending on whether the implicit variable lives on the dual/regular grid, respectively. 

Note that the above stencils do not include the contribution from the domain boundaries, which can be dealt with a corresponding boundary condition stencil for both divergence and gradient operators. For a divergence boundary condition, the following form is assumed

\begin{equation}
S_{ij}u_j = \pm  \delta_{ii_b} \frac{J_{i_b}F_{b}u_{b}}{V_{i_b}},
\end{equation}
where $i_b$ is the spatial index of the boundary cell (with $J_{i_b}$ denoting either the left or right face Jacobian in this case), and the sign depends on whether the boundary is the left(+) or right(-) domain boundary. $F_{b}$ is the flux Jacobian variable, and both it and $u$ are linearly extrapolated onto the boundary. This form assumes that the flux through the boundary is given as $F_bu_b$, with the flux Jacobian variable living on the regular grid. The boundary condition operator written in this form allows for the application of a lower bound to the flux Jacobian, which is useful, for example, in setting the Bohm condition at the divertor target (see \ref{appendix:SOL-SK}). For a gradient operator both the flux and face Jacobians are ignored, and the stencil becomes simply

\begin{equation}
S_{ij}u_j =  \pm \delta_{ii_b} \frac{u_{b}}{dx_{i_b}}.
\end{equation}
Note that the above divergence boundary term is non-linear, due to both $F$ and $u$ being variables. More involved stencils might have derivation rules associated with them, or might require access to model-bound variables. An example is the spatial diffusion stencil, which requires both the evolved and the implicit variable to live on the regular grid, and takes in a derivation rule in order to calculate the diffusion coefficient $D$

\begin{equation}
\label{eq:diff-stencil}
S_{ij}u_j =  \frac{1}{V_i}\left(J_iD_{i+1/2}\frac{u_{i+1}-u_i}{dx_{i+1/2}}-J_{i-1}D_{i-1/2}\frac{u_{i}-u_{i-1}}{dx_{i-1/2}}\right).
\end{equation}

Kinetic/distribution variable stencils tend to be even more involved, and will not be covered in detail in this manuscript. An example of this complexity is the velocity space derivative operator, representing the term

\begin{equation}
    \frac{\partial f_l}{\partial t} = \frac{\partial }{\partial v}(C(v) f_{l'}),
\end{equation}
where the function $C(v)$ can be specified in the code as either a fixed velocity space vector, or a model-bound single harmonic variable. Similarly, $f_{l'}$ can be interpolated onto the velocity space cell boundaries, and the linear interpolation coefficients can be specified in a similar way to $C(v)$, defaulting to interpolating directly onto the cell boundaries. An example where both custom interpolation and $C(v)$ are useful is for the Chang-Cooper-Langdon scheme\cite{Epperlein1994} for the Coulomb collision operator (used in the model in \ref{appendix:SOL-SK}).

Other kinetic stencils include 

\begin{itemize}
    \item A moment stencil - represents taking the $n$-th moment of a harmonic $4 \pi \int_0^\infty f_lv^{n+2}dv$.
    \item Velocity space diffusion - with the ability to specify single harmonic model-bound data diffusion coefficients.
    \item Logical boundary condition - a stencil representing the logical boundary condition, as derived for SOL-KiT\cite{Mijin2021}.
    \item Spatial difference stencil - $\partial / \partial x$ operator for harmonics, behaving either as a central difference or a staggered (forward/backward) difference stencil, depending on where the individual harmonics are defined.
    \item Boltzmann collision operator stencil - based on the SOL-KiT Boltzmann collision integral implementation and using the collisional-radiative model-bound data (see next section)
    \item Other niche stencils - such as a stencil taking the moment of a kinetic term, or stencils designed for particular Coulomb collision operator terms.
\end{itemize}

A particularly convenient stencil from a user's perspective is the custom 1D fluid stencil, which gives the high-level user effectively low-level access with a very flexible interface. This stencil is defined as

\begin{equation}
\label{eq:custom-fluid}
    S_{ij} = \sum_k \delta_{i,j+s_k}X_{k,i}v_{k,i}w_{k,i},
\end{equation}
where $s_k$ defines the $k$-th relative stencil entry position. $X_{k,i}$ is the $i$-th entry of the fixed stencil component for the $k$-th relative stencil entry, while $v_k$ and $w_k$ represent \textbf{VariableContainer} and \textbf{ModelboundData} fluid variables that can be included as individual stencil columns. In this way, combined with various \textbf{Derivations}, the user can represent most fluid stencils. An example would be a three point stencil, where $s = [\,-1 \quad 0 \quad  1\,]$, meaning that each spatial location requires information from its left neighbour, itself, and its right neighbour. This three point stencil can then be used, for example, to represent the diffusion stencil in equation (\ref{eq:diff-stencil}) by setting 

$$X_{1,i} = \frac{J_{i-1/2}}{V_{i}dx_{i-1/2}}, \quad X_{3,i} = \frac{J_{i+1/2}}{V_{i}dx_{i+1/2}}, \quad X_{2,i} = -X_{1,i}-X_{3,i},$$
where the diffusion coefficient was assumed to be 1, for simplicity\footnote{Otherwise there would be a need to define three derived variables to use as $v_k$ or $w_k$ in equation (\ref{eq:custom-fluid}), for which one could use calculation trees}. One could then add complexity by accounting for boundary conditions, for example by setting $X_{1,i}$ to 0 at the left boundary and $X_{3,i}$ to 0 at the right boundary. 

More details on the available stencils and their options can be found in the code documentation.

\subsection{Collisional-radiative model-bound data}

While one could write a collisional-radiative model (in the ODE sense from Section \ref{ProblemClasses}) purely using derived variables and diagonal stencils, these models tend to contain many transitions and terms, so the cognitive load on a user implementing them term-by-term would be high. When one factors in the special treatment needed for Boltzmann collisions, it becomes natural to group collisional-radiative data to allow for efficient \textbf{Term} generation. The \textbf{ModelboundCRMData} class serves this purpose, unifying \textbf{Transition} objects and inelastic collision mapping, and enabling a simple interface for \textbf{Term} generation. 

It is also worth noting here that species data in ReMKiT1D can be grouped into species objects, each being associated with a species name and integer ID, and containing data on the species charge and mass. Most importantly, it also allows for the association of certain variable names to a species, which further facilitates the automatic generation of \textbf{Terms}. The abstract \textbf{Transition} class assumes that each species is associated with an integer ID when defining the ingoing and outgoing states. For example, the integer ID 0 is always associated with electrons, and negative IDs are generally used for ion species, while positive IDs tend to denote neutrals species. As such, one could represent the ionization reaction 

$$ e^- + H \rightarrow  e^- + e^- + H^+,$$
as a \textbf{Transition} from ingoing to outgoing states/species\footnote{In the context of the CRM model-bound data these two terms are used interchangeably. It is assumed that each tracked state is represented by a species in the model.} $[\, 0 \quad  1\,] \rightarrow [\, 0 \quad 0 \quad  -1\,]$, assuming that the hydrogen atom is given the ID 1 and $H^+$ the ID $-1$. Then, depending on the concrete \textbf{Transition} class, the following quantities are accessible through the abstract interface:

\begin{itemize}
    \item The reaction rate as a function of position
    \item The momentum loss rate as a function of position - usually available only for a small subset of electron induced transitions
    \item The energy loss rate as a function of position - generally associated with the energy loss of electrons
    \item Cross-section - electron impact cross section associated with the transition and used when constructing Boltzmann collision operators (can be a function of position in the general case, see below for detailed balance transition)
    \item Transition energy - either a fixed value or a value for each spatial cell based on the ratio of energy loss and reaction rates
\end{itemize}
Different \textbf{Transition} classes are available to the user, allowing for control over how the above quantities are calculated and used. The following are available and widely used as of v1.0.x:

\begin{itemize}
    \item \textbf{SimpleTransition} - the simplest possible transition with a fixed transition energy and rate
    \item \textbf{DerivedTransition} - a transition with the reaction rate calculated using a \textbf{Derivation}. Uses a fixed transition energy by default, but can also associate \textbf{Derivations} with the momentum and energy loss rates
    \item \textbf{FixedECSTransition} - a transition with a fixed transition energy and cross-section (which can be specified for any number of electron distribution harmonics). The rates are calculated using the cross-section, together with inelastic grid mappings (see below)
    \item \textbf{DBTransition} - a transition obtained using detailed balance (see SOL-KiT paper\cite{Mijin2021}) with another transition which has a cross-section associated with it. The resulting cross-section held by this \textbf{Transition} object is thus also a function of position.
\end{itemize}
In order to use Boltzmann collision operators or any of the transitions with associated cross-sections, inelastic transition grid data must be generated, in the same way as in SOL-KiT, by specifying a set of transition energies. Then transitions such as the \textbf{FixedECSTransition} and stencils such as the Boltzmann collision operator stencil can calculate rates and operators that obey particle and energy conservation.
Term generator objects can be created in ReMKiT1D that can scan the \textbf{ModelboundCRMData} object. They then can generate all terms corresponding to the particle and/or energy loss/gain rates due to collisions, as well as any Boltzmann collision operators. For example, particle sources can be generated using the reaction rate data of each transition, multiplying them with the ingoing state densities corresponding to the transition\footnote{Some transition rates that use cross-sections already include one electron density factor, so these are automatically dropped by generators} and by the population change due to the transition. In the above electron impact ionization example the population change for electrons and ions is $+1$, while it is $-1$ for the atoms. In general, these produce terms of the form

\begin{equation}
    \left(\frac{\partial n_b}{\partial t}\right)_T = P_b^TK^T\prod_{b' \in I_T}n_{b'},
\end{equation}
where $b$ is the species index (and can be associated with any species in either the ingoing or outgoing state lists), $I_T$ is the ingoing state list for transition $T$, $K^T$ is the reaction rate of the transition, and $P_b^T$ is the population change of species $b$ in transition $T$.
These sorts of generated terms are diagonal stencil terms using model-bound rate data and are implicit in the final ingoing state density (given as the first associated variable for that species). Other term generators can be found in the code documentation and examples.

\section{Verification and benchmarking}
\label{Benchmarking}

In order to build confidence in the implementation of various operators both in the Fortran and the Python interfaces of ReMKiT1D a large number of tests have been performed. Some of those will be presented here, aiming to cover a wide range of use cases.

\begin{table}[]
    \begin{tabular}{@{}ll@{}}
    \toprule
    Name and Section                                                               & Scripts                                                                                                                                                                                                                                                                   \\ \midrule
    \begin{tabular}[c]{@{}l@{}}Fluid advection test\\ 5.1.1.\end{tabular}          & ReMKiT1D\_advection\_test.ipynb                                                                                                                                                                                                                                           \\ \midrule
    \begin{tabular}[c]{@{}l@{}}MMS test \\ 5.1.2.\end{tabular}                     & ReMKiT1D\_MMS.ipynb                                                                                                                                                                                                                                                       \\ \midrule
    \begin{tabular}[c]{@{}l@{}}Kinetic advection test\\ 5.2.1.\end{tabular}        & ReMKiT1D\_kin\_adv\_test.ipynb                                                                                                                                                                                                                                            \\ \midrule
    \begin{tabular}[c]{@{}l@{}}Coulomb collision operators\\ 5.2.2.\end{tabular}   & \begin{tabular}[c]{@{}l@{}}ReMKiT1D\_ee\_coll\_test.ipynb \\ ReMKiT1D\_ee\_coll\_test.ipynb \\ ReMKiT1D\_cold\_ion\_test.ipynb\end{tabular}                                                                                                                               \\ \midrule
    \begin{tabular}[c]{@{}l@{}}Epperlein-Short test\\ 5.2.3.\end{tabular}          & \begin{tabular}[c]{@{}l@{}}ReMKiT1D\_ES\_test.ipynb\\ es\_test.py\\ es\_verif.ipynb\end{tabular}                                                                                                                                                                          \\ \midrule
    \begin{tabular}[c]{@{}l@{}}Collisional-Radiative tests\\ 5.3.\end{tabular}     & \begin{tabular}[c]{@{}l@{}}ReMKiT1D\_crm\_example.ipynb\\ ReMKiT1D\_kin\_crm\_test.ipynb\end{tabular}                                                                                                                                                                     \\ \midrule
    \begin{tabular}[c]{@{}l@{}}Epperlein-Short scaling tests\\ 5.4.1.\end{tabular} & \begin{tabular}[c]{@{}l@{}}es\_test\_weak\_scaling.ipynb\\ es\_test\_strong\_scaling.ipynb\end{tabular}                                                                                                                                                                   \\ \midrule
    \begin{tabular}[c]{@{}l@{}}SOL-KiT-like scaling tests\\ 5.4.2.\end{tabular}    & \begin{tabular}[c]{@{}l@{}}sk\_comp\_thesis.py\\ sk\_comp\_thesis\_kin.py\\ ReMKiT1D\_SK\_comp\_staggered\_kin\_thesis.ipynb\\ ReMKiT1D\_SK\_comp\_staggered\_thesis.ipynb\\ sk\_comp\_thesis\_strong\_fluid.ipynb\\ sk\_comp\_thesis\_strong\_kinetic.ipynb\end{tabular} \\ \bottomrule
    \end{tabular}
    \caption{Script names associated with each reported benchmarking and scaling test. All .py and those .ipynb files starting with ReMKiT1D are part of the RMK\_support Python package's examples. All other .ipynb files are part of the supplemental material for this manuscript.}
    \label{tab:1}
    \end{table}

It should also be noted that the Fortran code for ReMKiT1D comes with its own unit test suite, built using the testing framework pFUnit\cite{pFUnit}, with those test integrated into the GitHub repository, while the Python package is tested using the pytest package. In general, the lowest level operators in Fortran, such as individual \textbf{Derivations} and stencils are subjected to unit tests within the Fortran codebase, while higher level I/O routines are mostly covered by integration tests such as the verification tests presented in this section\footnote{In the future, mocking of JSON configuration files could be used to produce unit tests for the Fortran I/O routines for each feature.}. Furthermore, several integration tests of common models are included in the unit test suite of the Fortran codebase. For more details on these tests see the corresponding repositories. 

Beyond benchmarking, a number of parallel performance/scaling tests have been conducted, and those will also be covered in this section. 

Table \ref{tab:1} lists all of the tests performed in upcoming sections, as well as the scripts associated with them, which are available either as parts of the Python package's examples or as supplemental materials for this manuscript.

\subsection{Verification of fluid operators}

A number of fluid operators are implemented in the code, as noted in Section \ref{Operators}. These are primarily divergence and gradient operators used to represent terms such as advective and pressure gradient terms, as shown in the Python example in the previous section. 

\subsubsection{Simple advection}

\begin{figure}[h!]
     \centering
     \begin{subfigure}[b]{0.45\textwidth}
         \centering
         \includegraphics[width=\textwidth]{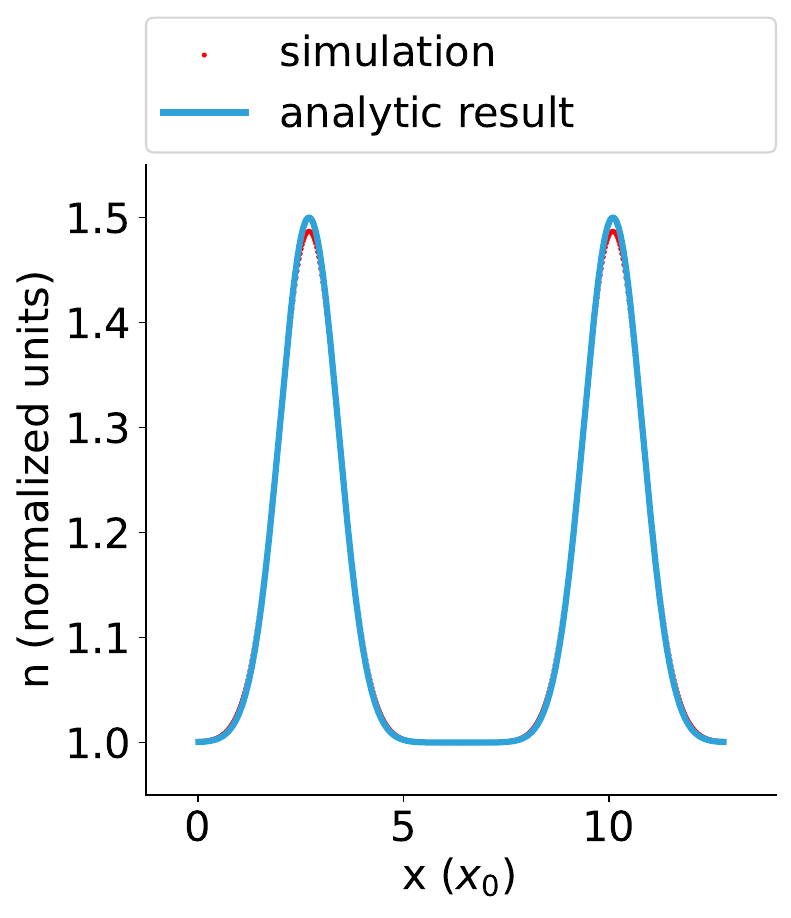}
         \caption{Comparison of analytic and numerical results at the end of the simulation $t=1000t_0$}
     \end{subfigure}
     \hfill
     \begin{subfigure}[b]{0.5\textwidth}
         \centering
         \includegraphics[width=\textwidth]{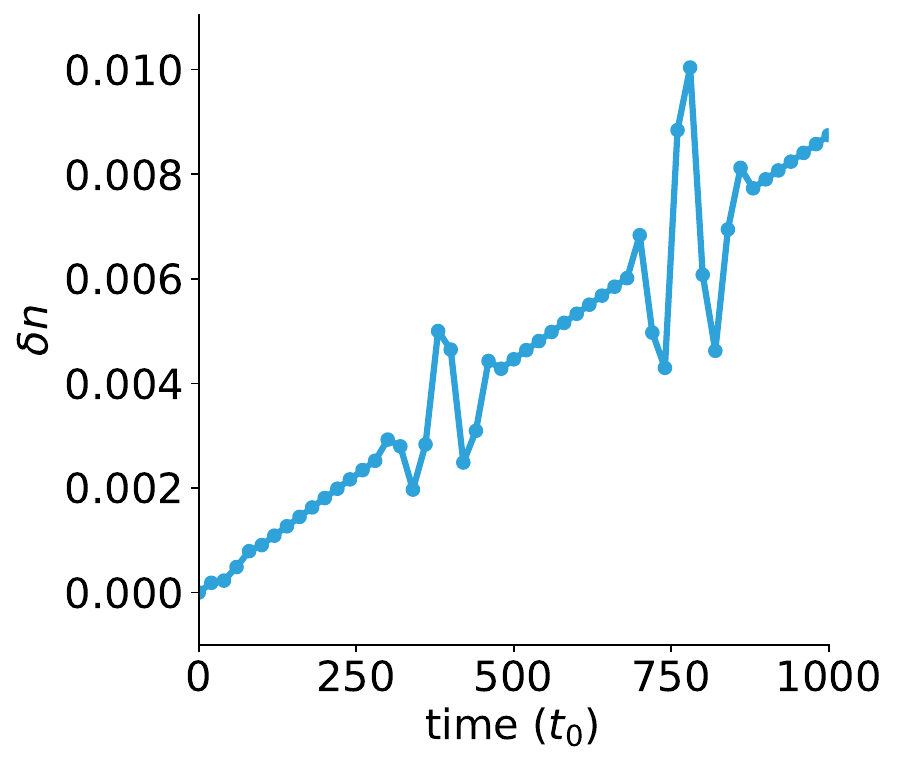}
         \caption{Maximum relative error of numerical result during simulation}
          \label{fig:adv_test_b}
     \end{subfigure}
        \caption{Advection test results compared with analytical solution}
        \label{fig:adv_test}
\end{figure}

Figure \ref{fig:adv_test} shows the result of running the advection test from the previous section for 1000 normalized times and comparing to the analytical result. Agreement is generally good, with the relative error during the simulation shown in Figure \ref{fig:adv_test_b}. The relative error is defined here as 

$$\delta n = \max(|n_{simulation} - n_{analytic}|/n_{analytic}),$$
where the maximum is taken along the spatial domain. Note that the two oscillatory features in the relative error in the figure come from the reflection of the wave at the boundaries.

Note that the above implementation of the advection operators does not include any flux limiting or artificial viscosity. ReMKiT1D v1.0.x does allow for the inclusion of artificial viscosity using the calculation tree approach. Future implementations of non-matrix terms will address more complex flux-limiter schemes used for shock capturing.

\subsubsection{MMS test of isothermal 2-fluid model}

In order to test a slightly more involved problem, the following equations were implemented:

\begin{align}
    &\frac{\partial n_s}{\partial t} + \frac{\partial \Gamma_s}{\partial x} = 0,\\
    &m_s\frac{\partial \Gamma_s}{\partial t} + \frac{\partial}{\partial x} \left(n_skT_s + m_s\Gamma_s u_s\right)- Z_s e n_s E = 0, \\
    &\frac{\partial E}{\partial t} = - \frac{j}{\epsilon_0},
\end{align}
where $s$ is a species index, here either for electrons or deuterium ions, and $j = \sum_s Z_s e \Gamma_s$ is the total current, making the electric field equation, when solved implicitly, act as a current constraint (see SOL-KiT implementation). $\Gamma_s = n_s u_s$ is the particle flux, and $Z_s$ is the species charge. The species temperatures $T_s$ are left as constants. These equations can be suitably normalized to be 

\begin{align}
    &\frac{\partial n_s}{\partial t} + \frac{\partial \Gamma_s}{\partial x} = 0,\\
    &\frac{\partial \Gamma_s}{\partial t} + \frac{\partial}{\partial x} \left(\frac{m_e}{2m_s} n_sT_s + \Gamma_s u_s\right)- \frac{m_e}{m_s}Z_s n_s E = 0, \\
    &\frac{\partial E}{\partial t} = - t_0^2\omega_p^2j,
\end{align}
where $t_0$ is the normalization/electron-ion collision time and $\omega_p$ is the plasma oscillation frequency at the normalized density. The normalized temperature $T_s$ is set to $0.5T_0$. In order to test the above equations, the following manufactured solution is used with reflective boundary conditions:

\begin{align}
    n_s &= 1 + 0.1 \frac{x-L}{L},\\
    u_s &= - 0.01 x \frac{x-L}{L^2},\\
    E &= - \frac{1}{4} \frac{1}{n_e} \frac{\partial n_e}{\partial x} -  \frac{1}{n_e} \frac{\partial}{\partial x} (nu^2) ,\\
\end{align}
where the fact that the temperature is equal to $0.5T_0$ is used explicitly and the electric field is calculated from the electron momentum equation. $L$ is the length of the domain, and $x$ the spatial coordinate, either on the regular or dual grid. Following the Method of Manufactured Solutions (MMS), the above solutions are inserted into the equations and the resulting source terms are added in order to push the solution towards the manufactured one. Note that the electric field equation is unaffected, as the manufactured solution assumes $j=0$. Finally, the densities $n_s$ are set to live on the regular grid, and the fluxes $\Gamma_s$ and electric field $E$ are set to live on the dual/staggered grid. The simulation is then run for several ($\approx 3$) sonic transition times $L/c_s$. $L$ here is set to $10\text{m}$ and the normalised sound speed is $c_s=\sqrt{m_eT_e/m_i}$.

\begin{figure}[h!]
     \centering
     \begin{subfigure}[b]{0.45\textwidth}
         \centering
         \includegraphics[width=\textwidth]{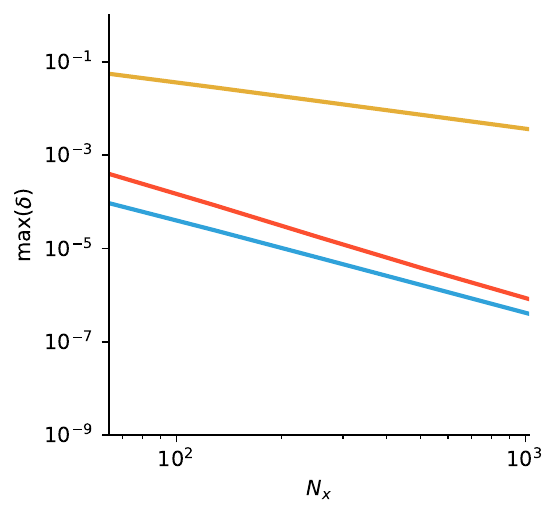}
         \caption{Without accounting for extended boundary cells in manufactured solution}
         \label{fig:mms_testa}
     \end{subfigure}
     \hfill
     \begin{subfigure}[b]{0.45\textwidth}
         \centering
         \includegraphics[width=\textwidth]{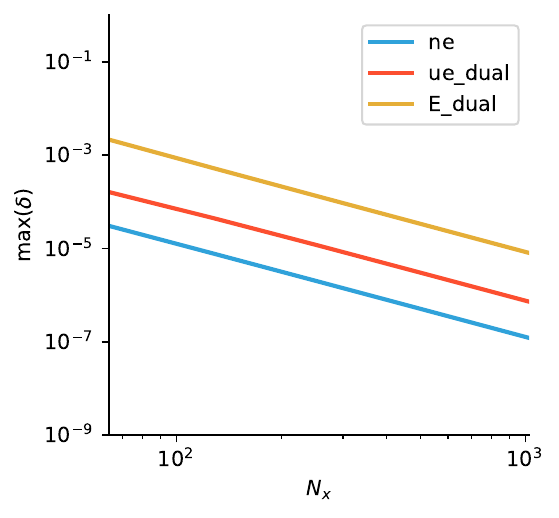}
         \caption{With accounting for extended boundary cells in manufactured solution}
         \label{fig:mms_testb}
     \end{subfigure}
        \caption{Convergence of MMS test on simple isothermal 2-fluid model with $N_x =64-1024$. The plotted quantity is the maximum relative error of the respective variables compared to the manufactured solution after $t\approx 3 L/c_s$}
        \label{fig:mms_test}
\end{figure}

The errors are calculated as the maximum (within the domain) relative departure of the tracked quantities compared to the initial values at the end of the simulation, and are shown in Figure \ref{fig:mms_test}. When the manufactured solution is computed directly, in particular the $\partial (\Gamma_s u_s)/\partial x$ terms, the electric field converges poorly due to discrepancies at the system boundaries, see \ref{fig:mms_testa}. This is because the default operators used in this example assume that the boundary cells on the dual grid are extended, as shown in Section \ref{Grids}. Once this is taken into account and those gradient terms are modified in the manufactured solution, much better spatial convergence is obtained, see \ref{fig:mms_testb}.
\subsection{Verification of kinetic operators}

A number of kinetic operators are included in ReMKiT1D, with some used to compose more complex operators, such as the Coulomb collision operators (see \ref{appendix:SOL-SK}). A number of these operators will be subjected to verification tests in this section.

\subsubsection{Kinetic advection}

\begin{figure}[h!]
    \centering
    \includegraphics[width=0.95\textwidth]{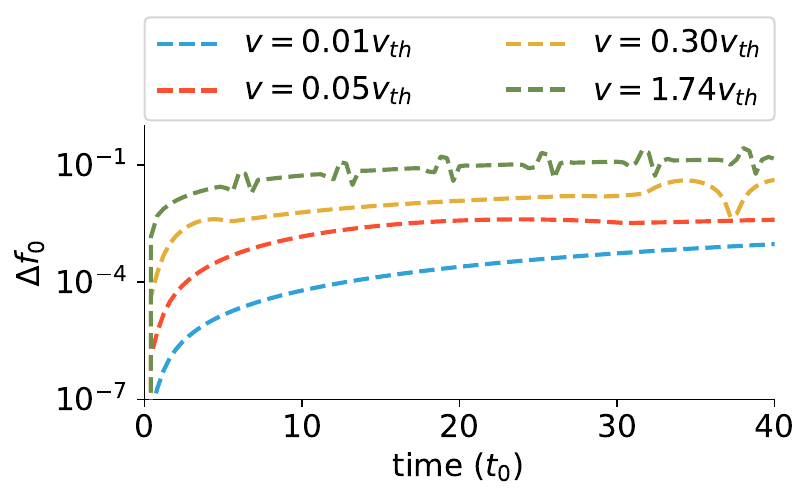}
    \caption{Absolute error of $f_0$ harmonic compared to analytical result due to spatial advection.}
    \label{fig:f0-adv}
\end{figure}

As noted in Section \ref{Grids}, on staggered grids the even harmonics live on the regular grid (cell centres) and the odd distribution harmonics live on the dual/staggered grid, so it is possible to write a simple advection test for the kinetic spatial advection operator that mimics the fluid advection setup by writing the equations for $f_0$ and $f_1$ without any fields or collisions

\begin{align}
    \frac{\partial f_0}{\partial t} + \frac{1}{3}v\frac{\partial f_1}{\partial x} = 0, \\
    \frac{\partial f_1}{\partial t} + v\frac{\partial f_0}{\partial x} = 0, 
\end{align}
which gives a wave equation for $f_0$ with wave speed $v/\sqrt{3}$. By initializing $f_0$ spatially as a Gaussian for all velocities $v$ one can then test the numerical errors for each velocity grid. These are, as expected from keeping the same time and space discretization, worse for larger velocities, as shown in Figure \ref{fig:f0-adv}. Similar to the fluid advection test the reflections from boundaries are seen as oscillations in the error. While Figure \ref{fig:f0-adv} shows a worryingly high error for high velocities, it should be noted that the Gaussian initial condition is the same for all velocities, so the distribution function is unphysically large at high velocities, where it would be orders of magnitude smaller than in the bulk, so in practice this error at high velocities contributes very little to the moments of the distribution function. 

For completeness, the grid parameters for this test are as follows. The spatial grid has normalized length $L=12.8x_0$ with 128 cells, and the simulation is run with time steps of $\Delta t = 0.01 t_0$, where we note again that $v_{th}=x_0/t_0$. In this example $v_{max}dt/dx = 1$, resolving all of the wave speeds in the system, albeit poorly for higher $v$ values, as evident from Figure \ref{fig:f0-adv}. For more details the reader is directed to the relevant example script.

\subsubsection{Coulomb collision operators}

Coulomb collision operators are an important part in accurately modelling electron kinetic effects, and the implementation based on ReMKiT1D velocity space operators will be tested in this section. For more details on the functional form and numerical implementation, the reader is directed to previous work in SOL-KiT, as well as \ref{appendix:SOL-SK} and the relevant Python routines referenced in the scripts. 

\begin{figure}[h!]
     \centering
     \begin{subfigure}[b]{0.45\textwidth}
         \centering
         \includegraphics[width=\textwidth]{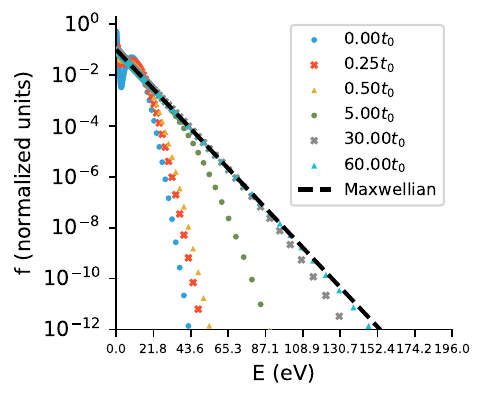}
         \caption{Relaxation of bump-on-tail distribution to a Maxwellian}
     \end{subfigure}
     \hfill
     \begin{subfigure}[b]{0.45\textwidth}
         \centering
         \includegraphics[width=\textwidth]{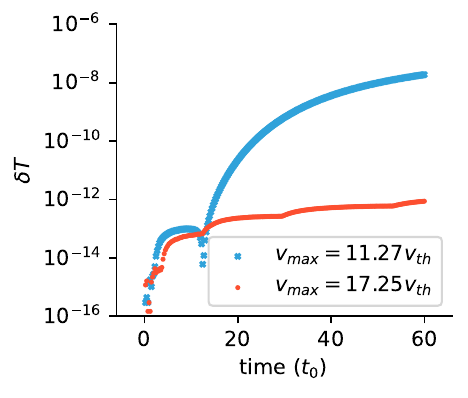}
         \caption{Temperature error during bump-on-tail relaxation for two different grid lengths with the same number of cells}
         \label{fig:f0-e-e_b}
     \end{subfigure}
        \caption{$f_0$ relaxation test under electron-electron collisions}
        \label{fig:f0-e-e}
\end{figure}

The first Coulomb collision operator to be tested is the isotropic electron-electron operator, a non-linear operator acting on $f_0$, conserving particles and energy and pushing the distribution towards a Maxwellian. The conservative finite difference implementation of this operator should conserve particles exactly and energy up to non-linear tolerance. To test this, only the $f_0$ harmonic is evolved, with the following bump-on-tail initial condition (in normalized quantities)

$$ f_0(t=0) = \frac{n}{(T\pi)^{3/2}} e^{-v^2/T} + \frac{0.1n}{(T\pi)^{3/2}}  e^{-(v-v_{bump})^2/T},$$
where $n = 1n_0$, $T=0.5T_0$, and $v_{bump}=3v_{th}$, with standard normalization. This leads to the relaxation of the bump towards a Maxwellian with effective temperature $T\approx6.07T_0$. This relaxation is plotted in Figure \ref{fig:f0-e-e} on a log scale, with the x-axis being the energy grid (simply $v^2$ with the standard normalization). The time step used is $\Delta t = 0.05 t_0$, and the simulations were ran up to $t=60t_0$.

The velocity grid cell widths are given by $\Delta v_1 = 0.01535 v_{th}$ and $\Delta v_i = c_v \Delta v_{i-1}$, with a total of 120 cells. The test was performed with two values of $c_v$, $1.025$ and $1.03$, which give total velocity grid lengths of approximately $11.27v_{th}$ and $17.25 v_{th}$, respectively. As shown in Figure \ref{fig:f0-e-e_b}, the shorter grid has a much worse energy conservation, given by the relative error of the effective temperature. This is because the temperature is considerably larger than the normalization temperature $T_0$, leading to an under-resolved Maxwellian tail in the latter half of the relaxation on the shorter grid. While this resolution effect is important when there is a substantial evolution in the distribution tail during the simulation, if the solution is already close to equilibrium the error is less pronounced. However, care should always be taken that high energy electrons in kinetic runs are adequately resolved.

\begin{figure}[h!]
     \centering
     \begin{subfigure}[b]{0.45\textwidth}
         \centering
         \includegraphics[width=\textwidth]{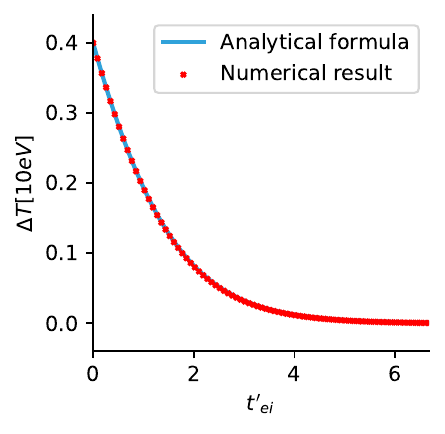}
         \caption{Comparison of e-i temperature relaxation to analytical result}
     \end{subfigure}
     \hfill
     \begin{subfigure}[b]{0.45\textwidth}
         \centering
         \includegraphics[width=\textwidth]{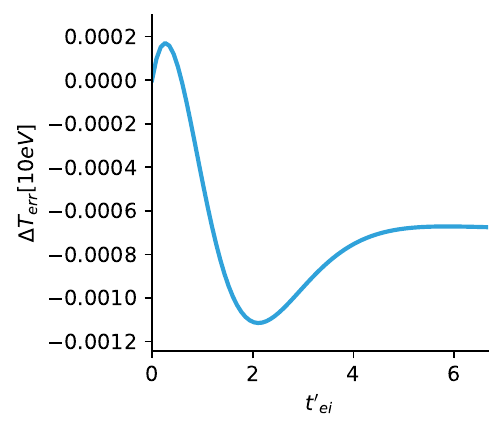}
         \caption{Over-relaxation in e-i collision test}
         \label{fig:f0-e-i-b}
     \end{subfigure}
        \caption{Temperature equilibration test under electron-ion collisions - due to finite grid effects}
        \label{fig:f0-e-i}
\end{figure}

The second collision operator to test is the electron-ion collision operator for $f_0$, which leads to temperature equilibration with fluid ions. To do this, both the collision operator and a term that takes its energy moment are added to the equations, to which an ion energy equation is now added, containing the moment term. For more information on the electron-ion operator, see \ref{appendix:SOL-SK} or references \cite{Power2021,shkarofsky1966particle}. To test the relaxation in the collisional limit, electron temperature is initialized at $T_e=8$eV and the ion temperature $T_i=4$eV, with both densities set to the normalization density $n_0=10^{19}\text{m}^{-3}$. In this case, the equilibrium temperature is $T_A=6$eV, and we expect both species to relax to that temperature. Following Shkarofsky\cite{shkarofsky1966particle}, we define 

\begin{align}
    \xi &= \frac{n_i\left(T_e-T_i\right)}{n_eT_e+n_iT_i},\\
    t'_{ei} &= \frac{8(n_e+n_i)}{3\sqrt{\pi}}\Gamma_{ei}\frac{m_e}{m_i}\left(\frac{m_e}{2kT_A}\right)^{3/2} t,
\end{align}
where $\Gamma_{ei}$ is the standard Coulomb collision constant. Taking the small mass ratio assumption, the relaxation follows the simple differential equation

\begin{equation}
    \frac{\partial \xi}{\partial t'_{ei}} = - \frac{\xi}{\left(1+\xi\right)^{3/2}},
\end{equation}
to which an analytical solution is readily obtained. This solution is compared to the numerical result obtained with ReMKiT1D in Figure \ref{fig:f0-e-i}, showing good agreement. The simulation is ran on the short grid from the previous electron-electron collision example and with time steps $\Delta t = 0.1t_0$. As can be seen from the absolute error of the solution in Figure \ref{fig:f0-e-i-b}, the final electron temperature is not exactly the same as the ion temperature, with the error well below 1\%. This is likely due to finite velocity grid effects.

\begin{figure}[h!]
    \centering
    \includegraphics[width=0.95\textwidth]{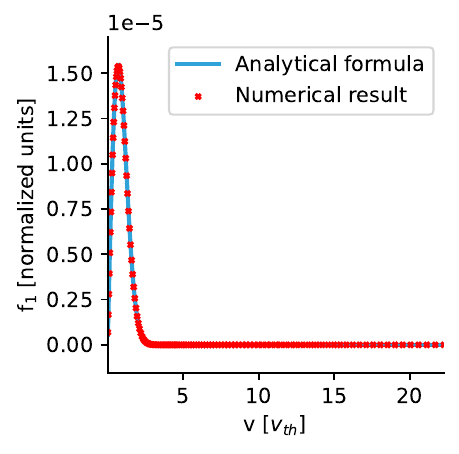}
    \caption{Comparison of numerical result of equilibrium $f_1$ harmonics with slowly drifting cold ions to analytical result.}
    \label{fig:f1-e-i}
\end{figure}

To test the electron-ion operator for $l=1$, responsible for momentum exchange, the following setup was used. The electron distribution function was initialized as a Maxwellian at the standard normalization density and temperature ($T_0=10$eV, $n_0=10^{19}\text{m}^{-3}$), with the ions initialized at the same density, and flowing at the speed $u_i=10^{-4}v_{th}$. The only terms solved were the cold ion electron-ion collision operator terms for $l=1$ (see \ref{appendix:SOL-SK} or SOL-KiT), as well as terms in the ion momentum equations that represent the friction moments of the collision operator terms. In the slow ion limit, the distribution function has the following solution for its $l=1$ harmonic

\begin{equation}
    f_1 = - u_i \frac{\partial f_0}{\partial v},
\end{equation}
which is readily computed for a Maxwellian $f_0$ and is compared to the numerical simulation in Figure \ref{fig:f1-e-i}. The total momentum is conserved up to solver tolerance, and the error in the equilibrium electron flow speed (expecting it to equal $u_i$) is 0.36\% on the same grid as the previous electron-ion $l=0$ operator test. 

Higher harmonic Coulomb collision operators are tested in the Epperlein-Short test to follow.

\subsubsection{Epperlein-Short test}

\begin{figure}[ht!]
    \centering
    \includegraphics[width=0.95\textwidth]{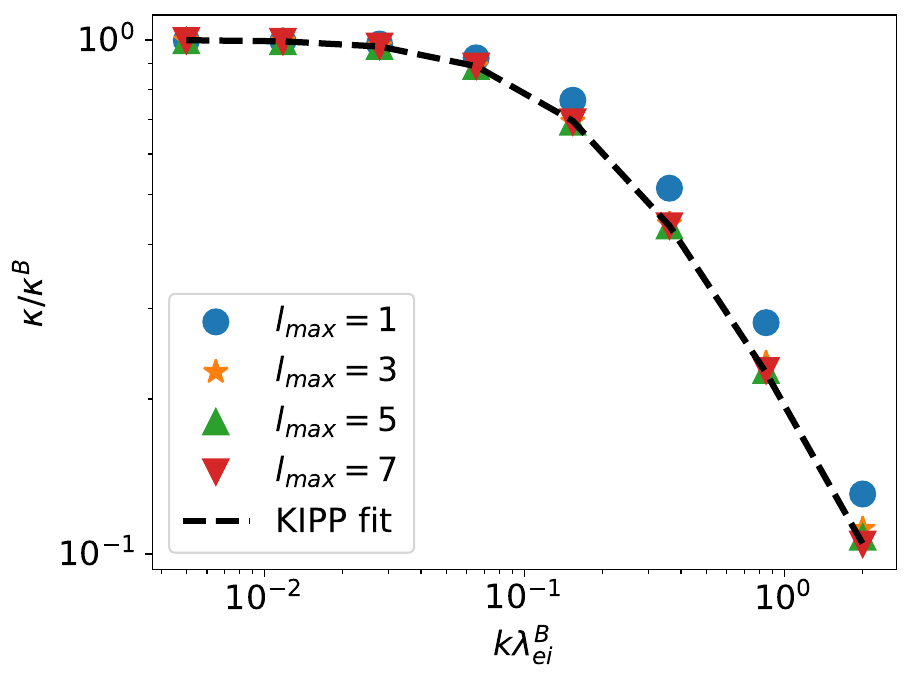}
    \caption{Convergence of ES test with number of harmonics. The fit is based on fit data from KIPP simulations published in \cite{Brodrick2017} and previously used to test SOL-KiT.}
    \label{fig:es_test}
\end{figure}

In order to fully test the electron-ion collision operators for higher harmonics the well-known Epperlein-Short (ES)\cite{Brodrick2017,Epperlein1990} test has been conducted. This entails a small electron temperature perturbation decay with electron-electron and electron-ion collisions enabled. Through the decay of the perturbation, the ratio of electron heat conductivity to the classical Braginskii\cite{Braginskii1965} value can be inferred, either through a direct comparison or through examining the decay rate. The initial conditions are set to 

$$T = T_0 + T_1 \sin(2\pi x /L),$$
where the perturbation wavelength can be controlled by modifying the domain length $L$. The used grid was periodic and contained $N_x=128$ spatial cells, $N_v=120$ velocity cells with widths given by $\Delta v_1 = 0.0307 v_{th}$ and $\Delta v_i = 1.025 \Delta v_{i-1}$. In this example, the normalization temperature was set to $T_0=100\text{eV}$, while other normalization quantities are set to the default values. Density is set to the normalization value, $T_1$ was set to $0.1\text{eV}$, and ion charge is left at 1. Following the approach used to benchmark SOL-KiT\cite{Mijin2021}, the ES test was performed for 4 values of $l_{max}$, and the results are plotted against the same fit based on KIPP\cite{Chankin2014,Brodrick2017} data in Figure \ref{fig:es_test}. Here $\lambda_{ei}^B$ can be converted to normalized length just as with SOL-KiT - $\lambda_{ei}^B=3\sqrt{\pi}/(4\sqrt{2})x_0$. Generally good agreement is found with previous SOL-KiT benchmarking, including the agreement with KIPP results, increasing confidence in the default collision operator implementation in the ReMKiT1D framework.

\subsection{Collisional-Radiative Model tests}

In order to test the CRM capabilities a number of tests have been run for both fluid and kinetic electrons. The implementation of inelastic Boltzmann collision operators is adapted from SOL-KiT, and the reader is encouraged to refer to the original SOL-KiT paper for the explanation of the particle and energy conserving discretization. For the tests shown here, the atomic data is in-built Janev\cite{Janev2003CollisionPI} hydrogen data\footnote{This data is hard-coded in the Fortran source code as an option, but the user is free to define their own data as well.}, together with NIST\cite{nist} spontaneous transmission rates. 

Two fluid and one kinetic electron test have been performed. The fluid tests were focused on detailed balance and the Saha-Boltzmann equilibrium, as well as a qualitative comparison with existing kinetic electron simulations performed by Colonna et al\cite{Colonna2001}. All tests are done in 0D (one spatial cell). The velocity space used to calculate Maxwellian moments of the Janev cross-sections in the fluid case is composed of $N_v=80$ cells with grid widths going from $\Delta v = 10^{-2} v_{th}$ to $v_{th}$ on a logarithmic grid. 

\begin{figure}[h!]
     \centering
     \begin{subfigure}[b]{0.45\textwidth}
         \centering
         \includegraphics[width=\textwidth]{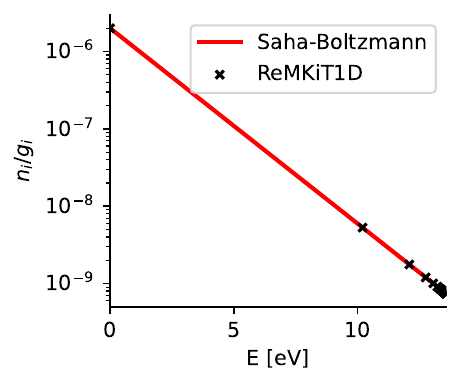}
         \caption{Saha-Boltzmann equilibrium with ReMKiT1D.}
         \label{fig:crm-fluid-a}
     \end{subfigure}
     \hfill
     \begin{subfigure}[b]{0.45\textwidth}
         \centering
         \includegraphics[width=\textwidth]{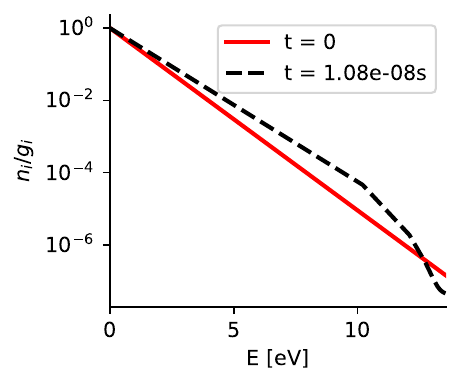}
         \caption{Evolution of atomic state distribution for fixed initial electron Maxwellian distribtution - to be compared with Figure 8 in \cite{Colonna2001}}.
         \label{fig:crm-fluid-b}
     \end{subfigure}
        \caption{Fluid tests of Collisional-Radiative Model with hydrogen. Here $n_i/g_i$ is the population of state with principle quantum number $i$ weighted by the state multiplicity ($g_i\propto i^2$ for hydrogen)}.
        \label{fig:crm-fluid}
\end{figure}

For all tests the following hydrogen reactions have been included:

\begin{itemize}
    \item Electron impact excitation and de-excitation
    \item Electron impact ionization
    \item Three-body recombination
\end{itemize}
with the evolution test in Figure \ref{fig:crm-fluid-b} also including radiative de-excitation and recombination. 

For the two fluid tests the number of neutral states tracked was set to 25, including the ground state, while the electron temperature was set to $T_e=1.72T_0=17.2\text{eV}$ (which corresponds to approximately 20000K). Default normalization was used in all tests. For the Saha-Boltzmann test the total density was set to $n=n_0=10^{19}\text{m}^{-3}$, with the initial atomic state distribution set to a Saha-Boltzmann distribution at half the electron temperature. The final atomic state distribution after a large number of time steps is shown in Figure \ref{fig:crm-fluid-a}, showing equilibration at the expected Saha-Boltzmann distribution with $T=T_e$.

\begin{figure}[ht!]
    \centering
    \includegraphics[width=0.95\textwidth]{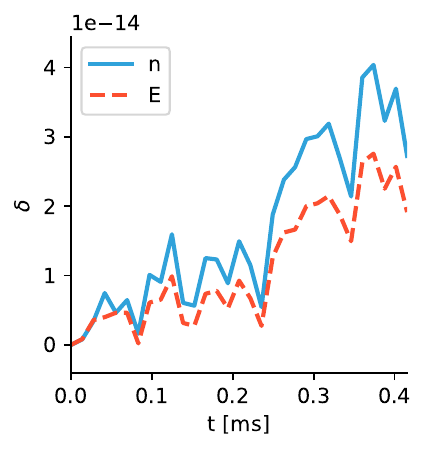}
    \caption{Conservation of heavy particles and total energy in kinetic electron CRM simulation with 20 neutral states.}
    \label{fig:crm-kinetic}
\end{figure}

The second test was the qualitative replication of the $t=10^{-8}s$ curve in Figure 8 of Colonna et al. For this purpose, the total density is initialized to $n=733893.9n_0$, loosely corresponding to 1 atmosphere of pressure at 1000K and with an initial ionization degree of $10^{-3}$. The electron temperature and initial atomic state distribution are set to the same as in the previous test. Finally, the electron distribution is clamped to its initial value. The result, shown in Figure \ref{fig:crm-fluid-b} qualitatively agrees well with the corresponding curve in Figure 8 in the original reference.
Finally, in order to test the conservation properties of the ReMKiT1D implementation of the SOL-KiT Boltzmann collision integrals for inelastic electron-neutral collisions, a long simulation was performed with 20 neutral states and with all non-radiative processes included. The velocity grid used had $N_v=120$ cells with widths given by $\Delta v_1 = 0.01 v_{th}$ and $\Delta v_i = 1.025 \Delta v_{i-1}$. The electron temperature was normalized to $T_0=5\text{eV}$ and its initial value was set to $T_0$. The initial electron density was set to $10^{19}\text{m}^{-3}$ and the initial neutral ground state density was set to $10^{18}\text{m}^{-3}$, with no excited state populations. Only inelastic collision integrals were included, allowing us to test the conservation properties isolating only these quantities. In order to isolate discretization errors from time integration errors, a low non-linear iteration relative tolerance of $10^{-14}$ was used. As shown in Figure \ref{fig:crm-kinetic}, both the energy and density relative errors are on the order of the iteration tolerance, even though the simulation was run for a macroscopically significant time and even though the total number of collision operators solved was above 400. 

\subsection{Parallel performance benchmarking}

In order to test parallel performance scaling a number of tests have been performed, mostly focusing on scaling in runs with kinetic electrons, as those are both generally more expensive and conducive to scaling, as well as allowing us to test the scaling behaviour of parallelization in the harmonic direction. All performance scaling tests have been done on the ARCHER2 HPC machine.

\subsubsection{Epperlein-Short test - strong and weak scaling}

A variant of the ES test used to verify collision operators has been used to test both strong and weak scaling, as well as investigate basic properties of harmonic parallelization available in ReMKiT1D. 

\begin{figure}[h!]
     \centering
     \begin{subfigure}[b]{0.45\textwidth}
         \centering
         \includegraphics[width=\textwidth]{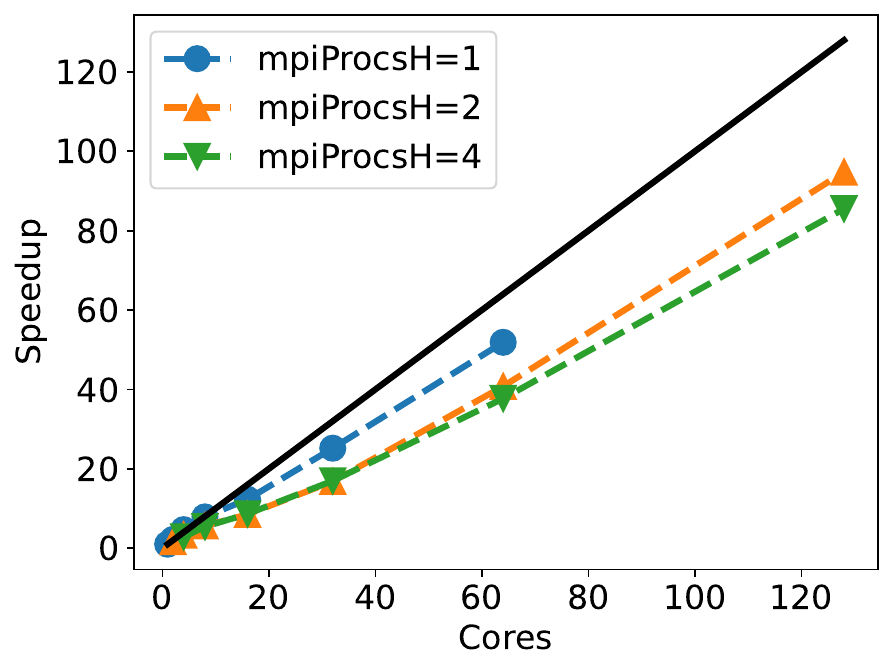}
         \caption{Strong scaling}
         \label{fig:es-scaling-a}
     \end{subfigure}
     \hfill
     \begin{subfigure}[b]{0.45\textwidth}
         \centering
         \includegraphics[width=\textwidth]{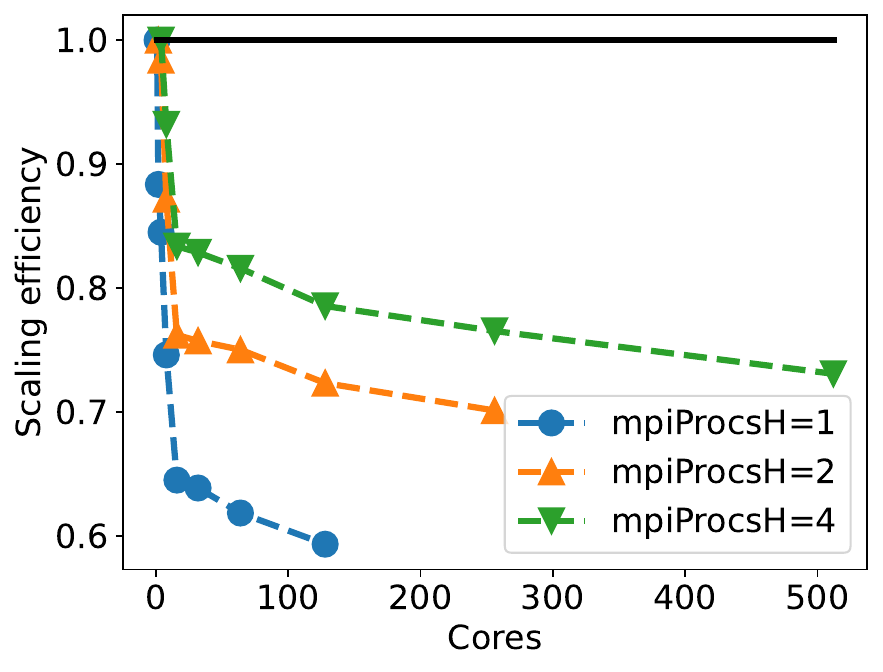}
         \caption{Weak scaling}
         \label{fig:es-scaling-b}
     \end{subfigure}
        \caption{Strong and weak scaling for the ES test. The different colours/markers represent different numbers of processes in the harmonic direction.}
        \label{fig:es-scaling}
\end{figure}

For the strong scaling, the following parameters were used. The maximum harmonic number was set to $l_{max}=7$, with $N_x=128$ spatial cells with width $\Delta x = 0.1x_0$, and the simulations were run for $N_t$ time steps with length $\Delta t = 0.05t_0$. Other parameters were set to the same in the verification test. Due to different behaviour of spatial and harmonic parallelization, strong scaling was tested for 1,2, and 4 processes in the harmonic direction. The results are shown in Figure \ref{fig:es-scaling-a}, where it can be seen that the runs with more harmonic direction processes scale up to a higher number of cores. However, it should be noted here that adding processors in the harmonic direction by default produces a more difficult matrix solve problem due to higher harmonics having shorter timescales. This leads to sub-matrices belonging to processes responsible for high harmonic number naturally being stiffer, leading to the 4 harmonic direction processor run with 256 cores failing due to the solver. While this could potentially be solved by introducing more involved preconditioning, that is beyond the scope of the present manuscript.

\begin{figure}[ht!]
    \centering
    \includegraphics[width=0.95\textwidth]{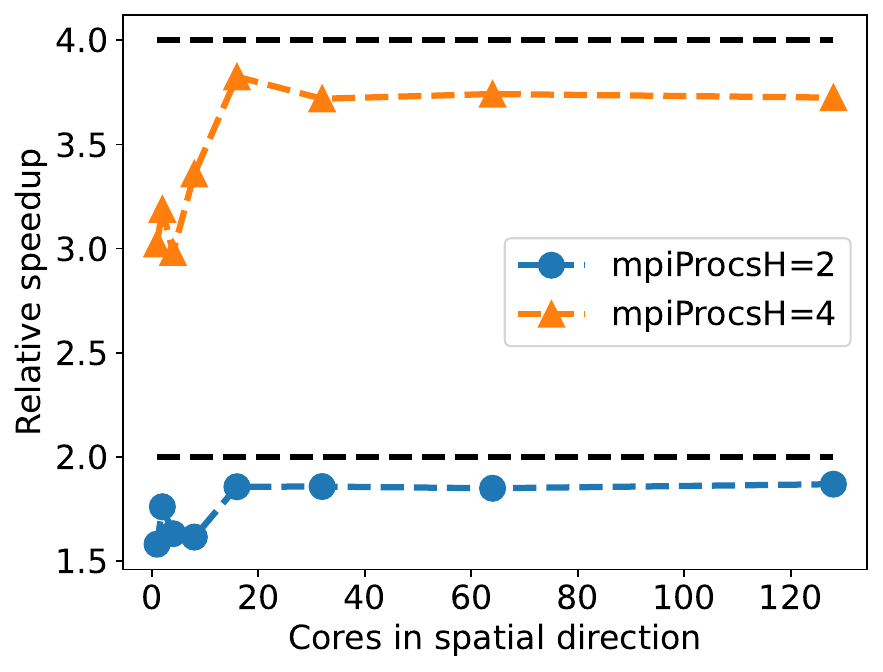}
    \caption{Relative speedup in the weak scaling set of runs from the ES test when processors are added in the harmonic direction. The black dashed lines are ideal speedups.}
    \label{fig:es-ws-speedup}
\end{figure}

For the weak scaling test, the problem was modified in order to avoid limitations due to the matrix solver. The total number of harmonics remains 8, but the length of the domain has been increased to $L=80x_0$, the time step length reduced to $0.001t_0$ and the number of time steps increased to $N_t=30000$. This way the main cost in the code was not the matrix solve, but the matrix construction. Scaling has been tested up to 4 ARCHER2 nodes, totalling 512 cores. The number of spatial cells was varied from 8 to 1024 in powers of 2, and the results are shown in Figure \ref{fig:es-scaling-b} for 1,2, and 4 processes in the harmonics direction. What can be seen is that in this example, where the code spends more resources on matrix construction instead of the solve, adding processors in the harmonic direction always improves performance. The relative speedup from adding harmonics is shown in Figure \ref{fig:es-ws-speedup}, where it can be seen it is close to the ideal speedup, particularly for higher numbers of processes in the spatial direction. 

\subsubsection{SOL models - strong scaling}

The final set of scaling tests is focused on testing more production-relevant models of the Scrape-Off Layer. Details of the models are given in \ref{appendix:SOL-SK}, and they will only briefly be summarized here. The models are loosely based on equations previously used in SOL-KiT, with the major difference being the use of the AMJUEL\cite{amjuel} database for effective hydrogen ionization and recombination. The electrons are treated either as a fluid or kinetically, and both options are tested here for scaling. The domain is reflective at the left boundary and has a sheath boundary condition at the right boundary, with a total domain length of $L=10\text{m}$, with the spatial grid being finer closer to the boundary. An effective heat flux of $3\text{MW/m}^2$ is injected over $L_h=3\text{m}$ upstream, while the ion temperature is assumed equal to the electron temperature. Standard normalization is used. The initial condition is based on a Two-Point Model solution\cite{Stangeby2000}, with the electron temperature upstream set to $T_u=20\text{eV}$ and downstream to $T_d=5\text{eV}$, and the upstream density set to $n_u=8\cdot10^{18} \text{m}^{-3}$. Initial conditions assume no neutral particles, which are injected through the recycling flux at the target. Neutrals are diffusive, while the ion continuity and momentum equations are explicitly solved, including charge-exchange interactions with the neutrals.

\begin{figure}[h!]
    \centering
    \includegraphics[width=0.95\textwidth]{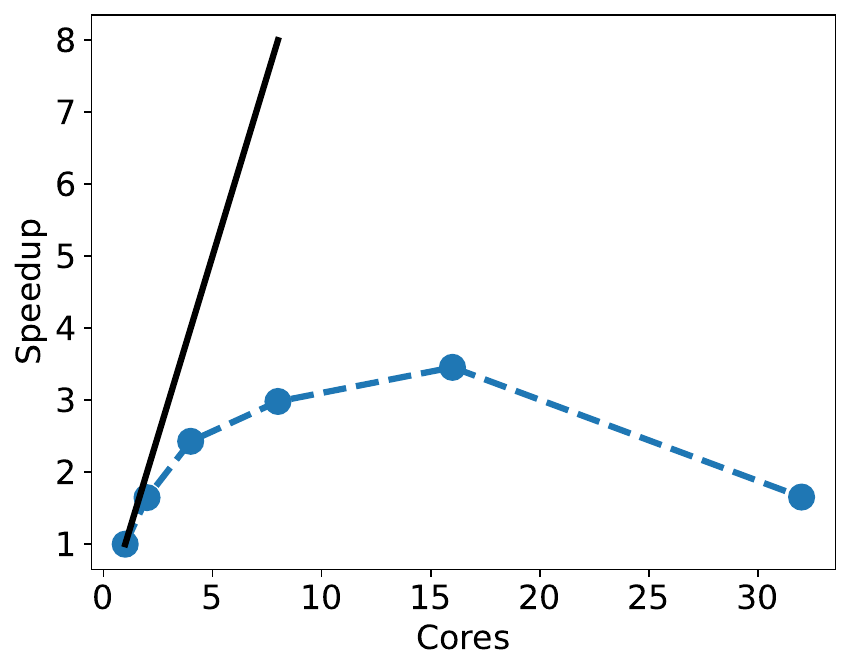}
    \caption{Strong scaling with a fluid electron SOL model showing poor scaling due to low number of degrees of freedom per processor}
    \label{fig:ss-sk}
\end{figure}

The fluid runs are set up with $N_x=512$ spatial cells and run until a time $t=9000t_0$, with time steps adaptively set to approximately 10\% of the shortest electron-ion collision time in the domain. Figure \ref{fig:ss-sk} shows the result of the scaling test up to 32 cores for this fluid problem. The scaling falls off very quickly due to limitations in the solver, with the simple explanation being that there are not enough degrees of freedom per core for the matrix solver to scale properly. 

\begin{figure}[h!]
     \centering
     \begin{subfigure}[b]{0.45\textwidth}
         \centering
         \includegraphics[width=\textwidth]{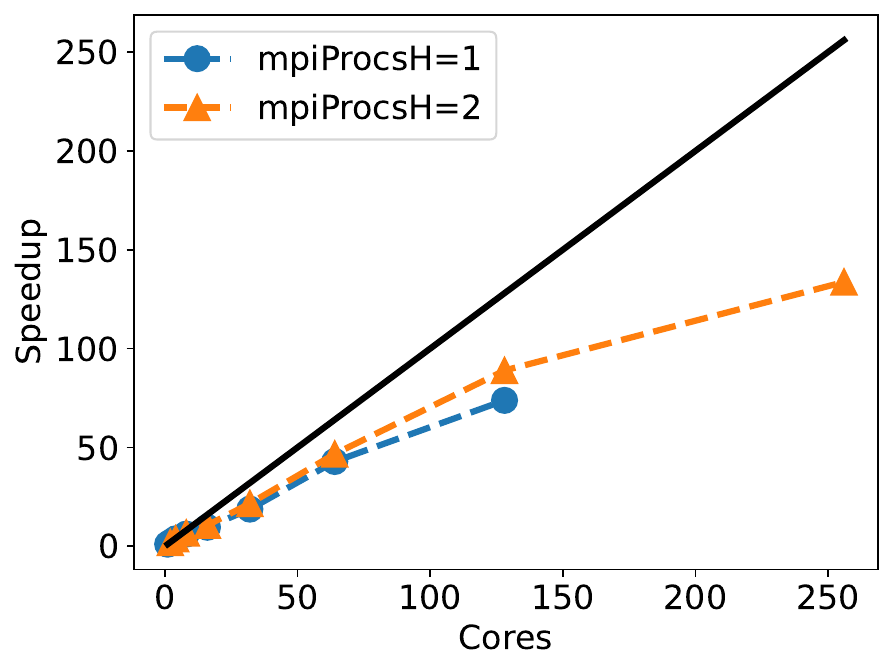}
         \caption{256 spatial cells and 4 harmonics starting with single core}
         \label{fig:ss-kin-scaling-a}
     \end{subfigure}
     \hfill
     \begin{subfigure}[b]{0.45\textwidth}
         \centering
         \includegraphics[width=\textwidth]{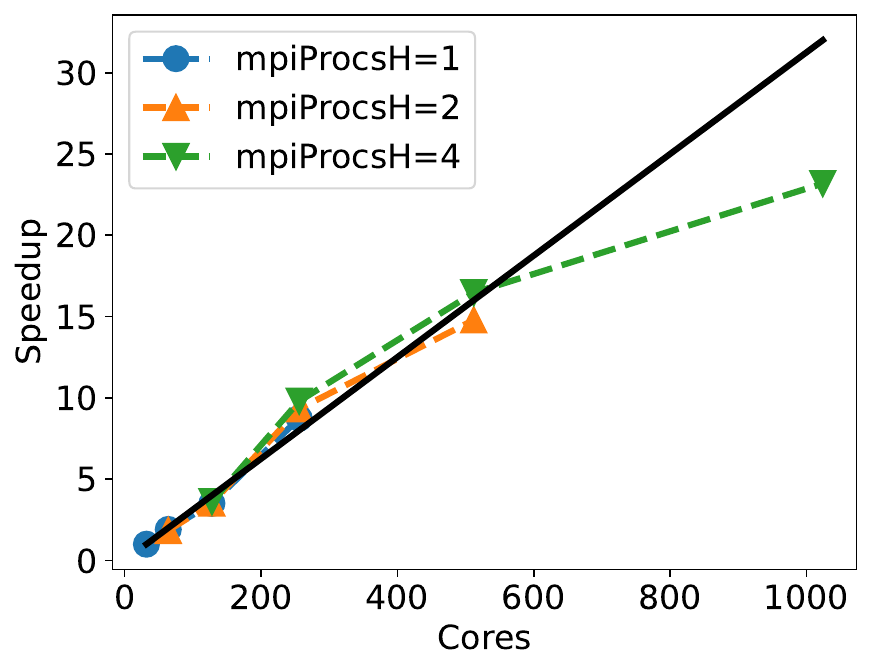}
         \caption{512 spatial cells and 8 harmonics starting with 32 cores}
         \label{fig:ss-kin-scaling-b}
     \end{subfigure}
        \caption{Strong scaling with a kinetic electron SOL model}
        \label{fig:ss-kin-scaling}
\end{figure}

For the kinetic electron test, two sets of runs were performed, a small scale set with $N_x=256$ and $l_{max}=3$ going from a single to 256 MPI processes, and a larger scale set with $N_x=512$ and $l_{max}=7$ going from 32 to 1024 MPI processes. Both sets were run up to $t=50t_0$ with time steps adaptively set to 5\% of the shortest electron-ion collision time. $N_v=80$ velocity cells are used, with cell widths ranging from $0.05v_{th}$ to $0.4v_{th}$. The results of these scaling tests are shown in Figure \ref{fig:ss-kin-scaling}. Unlike the Epperlein-Short test, adding processors in the harmonic direction for both sets of runs improves the scaling. This is likely due to this example having a more involved set of collision operators (due to the flowing ions), which shifts the cost of one solver iteration towards matrix building and away from the actual PETSc solver call. While Figure \ref{fig:ss-kin-scaling-b} seems to suggest a better-than-ideal speedup at first glance, this is simply due to the fact that that set of runs did not go down to serial due to the increased computational cost with 8 harmonics and 512 spatial cells, and the scaling seems to improve in the 64-512 core range compared to the 1-64 range. 

The results presented above showcase the increased scalability of kinetic models compared to fluid electron models, especially with the novel harmonic dimension domain decomposition. In the example above, moving from fluid to kinetic, the number of degrees of freedom associated with implicit electron variables goes from 5 per spatial cell to 320-640 per spatial cell (4-8 harmonics with 80 velocity space cells).
\section{Discussion}
\label{Discussion}
In this section the present limitations of the framework as well as potential use cases and future extensions will briefly be discussed. The focus will both be on the software design and numerical aspects, as well as on the achievable modelling with the current version of the software.

The main limitation in ReMKiT1D is its dimensionality, and the 1D aspect is baked into many parts of the framework. Another limitation is the hard-coded assumption that the distribution function is represented in a Legendre/Spherical Harmonic basis. However, basic conceptualization work is planned to explore the applicability of the Modeller-Model-Manipulator (3M) pattern in a way that is agnostic to both numerical methods and problem dimension. This would allow for solving the above two main limitations of the framework. 

Even in 1D, the framework's main strength is its flexibility, allowing for rapid iteration on models in the SOL. Examples of planned or ongoing applications include:

\begin{itemize}
    \item Equilibrium and transient simulations of the SOL, akin to those previously performed using SOL-KiT\cite{Mijin2020a}, but with flexible neutral and plasma models, as well as with multiple ion species, focusing on impurity transport and electron kinetic effects
    \item Exploration of different effective collisional operators that could be used in conjunction with both external and internal Collisional-Radiative Models to include impurity collisional effects on the electron distribution function in runs with kinetic electrons - this includes both simplified and high fidelity molecular deuterium effects
    \item The calibration of reduced models of SOL equilibria and transients against higher fidelity models both within ReMKiT1D and in other codes
    \item Training data production for machine-learning applications in the SOL
\end{itemize}
In order to properly handle some of the above applications, extensions to the framework might be required. Some of the extensions being planned or considered as options are:

\begin{itemize}
    \item Full support for explicit \textbf{Term} objects, as well as improved options for explicit time-stepping
    \item Adaptive Boltzmann collision operator stencils that can handle reactions with varying energy costs, such as those arising from CRMs
    \item Full support of Python level custom stencils for kinetic operators, allowing users to generate their own stencils in velocity space (the architecture behind the custom fluid stencil can be adapted for this)
    \item Multi-linear interpolation support for atomic data for use in the in-built CRM model-bound data
\end{itemize}

While flexibility and design scalability are the main priorities in ReMKiT1D development, performance represents an increasingly important aspect, particularly since multi-node scalability has been demonstrated for kinetic electron simulations. In order to improve performance, the coupling with PETSc should be explored in more detail, particularly in terms of preconditioners, while PETSc's GPU support might also be an option for more demanding kinetic runs in the future. Other improvements in performance could be bundled with the generalization of the 3M pattern, with a future version of the code using more efficient data structures in terms of memory access. 

Finally, the framework's Python interface is under active development with the aim to improve user-friendliness and introduce various quality-of-life features.

\section{Conclusion}
\label{Conclustion}

We have presented the new ReMKiT1D framework for the construction of 1D models of the tokamak Scrape-Off Layer(SOL). ReMKiT1D allows for the solution of a wide range of differential equations, particularly relevant to SOL modelling. In particular, ReMKiT1D enables the construction of coupled multi-fluid plasma models with a kinetic treatment of electrons, as well as a built-in capability to handle complex collisional-radiative models. While the emphasis in this manuscript has been on SOL modelling, there are no conceptual barriers to applying the capabilities demonstrated here to wider problems that fit within the classes of differential equations supported by ReMKiT1D. It should be noted, however, that there are likely many more optimization opportunities that should be exploited should the ReMKiT1D design start being used outside of the SOL modelling context. There are also potentially missing features, albeit some of them might be planned additions, such as shock-capturing methods.

One of the goals of the ReMKiT1D framework is a combination of flexibility and user-friendliness. On the path to that goal, we have used a combination of Object-Oriented Modern Fortran and high-level Python interfaces. The former enables reusable and performant abstractions, as well as coupling to established libraries such as PETSc, while the latter enables efficient high-level control of the framework in a widely used modern language. A human-readable JSON configuration file acts as a glue between the Fortran and Python parts of the framework. The Python library offers not only tools for the specification of runs, but also for the processing of the output data, which is in HDF5 format. A complete setup workflow using the Python library is presented after the basic concepts are introduced. This example, as well as the many scripts available on the Python repository, are meant to serve partly as an introductory tutorial.

We lay out the design philosophy of ReMKiT1D, which is embodied in the Modeller-Model-Manipulator (3M) pattern. The central part of the 3M pattern is the Modeller object, responsible for the centralization of integration calls and for storing variable data. It contains Models and Manipulators, to which it delegates the responsibilities of the representation of terms in the equations being solved, and the modification of variable data based on those representations, respectively. The strategy and composite patterns are widely used within the codebase. Manipulators represent composable strategies for modifying data, particularly in the context of time integration. Likewise, the concept of derivations, which encapsulate generalized function wrappers, enables flexible calculation of variables outside of the context of time integration. In particular, ReMKiT1D offers a way of transforming basic Python mathematical expressions into objects that can be injected into Fortran through the use of expression tree derivations.

In terms of numerical algorithms, ReMKiT1D currently offers implicit time integration, which is essential in many stiff problems arising in reactive plasma physics, as well as an explicit Runge-Kutta solver. The flexible design of the integrators allows for natural implementations of operator-splitting methods at the Python level, as well as multistep integration algorithms. In order to support the implicit solver natively, ReMKiT1D currently offers only matrix represenations of terms in the differential equations. Work is ongoing on supporting general terms, enabled by the flexibility behind the derivation system within the framework.

Various verification tests are presented, together with parallel performance tests, which are focused on the novel parallelization in distribution function harmonics. It is shown that this approach works and provides a significant improvement in the scalability of the framework when handling kinetic models. In particular, we show strong scaling of kinetic simulations up to 1024 cores (8 nodes) on ARCHER2, enabled by the partitioning in the harmonic dimension. As for the verification tests, both individual operator tests as well as standard integrated tests such as the method of manufactured solutions or the Epperlein-Short test have been performed, showing expected agreement of the implemented models.

Finally, ongoing and potential uses as well as future extensions of the framework are discussed in detail, focusing both on the software engineering aspects as well as model development using the framework. 

\section*{Declaration of competing interest} 

The authors declare that they have no known competing financial interests or personal relationships that could have appeared to influence the work reported in this paper.

\section*{Acknowledgments} 

We would like to thank the anonymous reviewers for helping us make improvements in the structure and presentation of this manuscript.

This work was granted access to the HPC resources of ARCHER2 by the Plasma HEC Consortium [grant number EP/X035336/1]. This work has been part-funded by the EPSRC Energy Programme [grant number EP/W006839/1].  To obtain further information on the data and models underlying this paper please contact PublicationsManager@ukaea.uk.  

For the purpose of open access, the authors have applied a Creative Commons Attribution (CC BY) licence (where permitted by UKRI, ‘Open Government Licence’ or ‘Creative Commons Attribution No-derivatives (CC BY-ND) licence’ may be stated instead) to any Author Accepted Manuscript version arising.

\appendix

\section{The SOL-KiT-like SOL models}
\label{appendix:SOL-SK}

Some modifications have been performed on the previously reported standard SOL-KiT model\cite{Mijin2021,Mijin2020a} for the purpose of testing the implementation in ReMKiT1D, with one major modification being the use of AMJUEL\cite{amjuel} rates instead of the SOL-KiT-style embedded CRM in order to reduce computational costs significantly. The equations for both the fluid models and the kinetic electron model will be presented in this appendix for completeness, while the reader is directed to previous SOL-KiT publications for more details.

\subsection{Fluid equations}

A minor difference between the SOL-KiT equations in previous publications and the equations implemented in the ReMKiT1D SOL-KiT-like models is that the fluid equations in ReMKiT1D's implementation are in conservative form, utilizing the capability to implicitly calculate variables with no explicit time derivative in their equations to extract the temperatures and heat fluxes in a way that keeps implicit stability. 

The electron fluid equations are given by:

\begin{align}
    &\frac{\partial n_e}{\partial t} + \frac{\partial \Gamma_e}{\partial x} = S,\\
    &m_e\frac{\partial \Gamma_e}{\partial t} + \frac{\partial}{\partial x} \left(n_ekT_e + m_e\Gamma_e u_e\right) + n_e e E = R_{ei}, \\
    &\frac{\partial W_e}{\partial t} + \frac{\partial}{\partial x} \left[\left(W_e + n_ekT_e\right)u_e + q_e\right] + \Gamma_e e E = Q_e, 
\end{align}
where $\Gamma_e=n_eu_e$ and $W_e = 3n_ekT_e/2 + n_e m_e u_e^2/2$. In order to facilitate future inclusions of multiple ion species the parallel transport coefficients for $q_e=\kappa_e\nabla kT_e$ and $R_{ei} = - 0.71n_e\nabla kT_e$ are calculated taking the Braginskii limit\cite{Braginskii1965}($m_e/m_i \rightarrow 0$ and $Z=1$) using expressions from Makarov et al\cite{Makarov2021}. The particle source $S$ results solely from ionization and recombination, and $Q_e=Q_h+Q_{en}$, where $Q_h$ is the upstream heating term, and $Q_{en}$ is the effective electron energy loss/source associated with ionization and recombination collisions. These are implemented using AMJUEL\cite{amjuel} rates H.4-2.1.5 and H.4-2.1.8 for the particle sources, and H.10-2.1.5 and H.10-2.1.8, together with the potential energy accounting for recombination (H.4-2.1.8. with 13.6eV), are used for $Q_{en}$. A specialized polynomial derivation is implemented in the framework to handle fits such as those in the AMJUEL database.

Note that $T_e$ and $q_e$ are actually treated as implicit variables using ReMKiT1D's capability to include temporally stationary variables in the implicit vector. This ensures stability due to the implicit nature of the scheme being kept, even if the plasma equations are solved in conservative form. 

Ion equations are given as 

\begin{align}
    &\frac{\partial n_i}{\partial t} + \frac{\partial \Gamma_i}{\partial x} = S,\\
    &m_i\frac{\partial \Gamma_i}{\partial t} + \frac{\partial}{\partial x} \left(n_ikT_e + m_e\Gamma_e u_e\right) - n_ie E = - R_{ei} + R_{CX}, 
\end{align}
where the assumption $T_i=T_e$ is taken as in the standard SOL-KiT model, and $R_{CX}=-\Gamma_in_nK_{CX}$ is the charge-exchange friction, with $K_{CX}$ using AMJUEL rate H.2-3.1.8, scaling the temperature dependence by $1/2$ to account for tracking deuterium instead of hydrogen.

The neutrals are diffusive with their density evolved according to 

\begin{equation}
    \frac{\partial n_n}{\partial t} = \frac{\partial}{\partial x}\left(D_n\frac{\partial n_n}{\partial x}\right) - S,
\end{equation}
where the diffusion coefficient is set to $D_n = k\sqrt{T_n T_e}/(m_iK_{CX}n_i)$, with the neutrals assumed to have $T_n=3\text{eV}$, corresponding to the Franck-Condon dissociation energy. The $\sqrt{T_e}$ factor comes from the ion thermal velocity, leading to the neutrals effectively having a diffusive temperature corresponding to a geometric mean between the Franck-Condon and ion temperatures.

Boundary conditions are set to reflective upstream, and sheath boundary conditions at the target, with $u_e=u_i=c_s=\sqrt{2kT_e/m_i}$ set by the Bohm condition and $q_e=\gamma_e \Gamma_e kT_e$, where the electron sheath heat transmission coefficient is approximately 4.86. Ions and electrons leaving the plasma are recombined on the surface and returned with 100\% recycling. Finally, the electric field is solved for by implicitly solving 

\begin{equation}
    \frac{\partial E}{\partial t} = - \frac{j}{\epsilon_0},
\end{equation}
where $j=e(\Gamma_i-\Gamma_e)$, which, together with the two momentum equations acts like a current constraint and ensures quasi-neutrality. 

\subsection{Electron kinetic equations}

When electrons are treated kinetically the electron kinetic equations are solved for the evolution of a set number of distribution function harmonics instead of the three electron moment equations in the previous section. The equations have been discussed in detail in the context of SOL-KiT\cite{Mijin2021,Mijin2020a}, and will not be repeated here for the sake of brevity. A brief description of the terms that remain unchanged from the SOL-KiT implementation will be given, with the effective cooling operators for use with AMJUEL rates introduced in more detail. 

The terms implemented from SOL-KiT are:

\begin{itemize}
    \item Spatial and velocity space advection (Vlasov) terms
    \item Coulomb collision terms for electron-electron collisions for all harmonics\footnote{Here the ability to define single harmonic variables in model-bound data comes in handy to implement Chang-Cooper-Langdon terms\cite{Epperlein1994}}
    \item Coulomb collision terms for electron-ion collisions for cold flowing ions for harmonics with $l>0$
    \item The logical boundary condition\cite{Mijin2021,Procassini1992} at the sheath using the previously developed harmonic formulation
    \item Diffusive heating terms upstream
    \item Secondary electron source/sink at low electron energy due to ionization/recombination
    \item Terms coupling the kinetic operators with the ion fluid equation through taking moments (e.g. $R_{ei}$) 
\end{itemize}

The only missing effect is the electron energy loss/gain terms due to ionization/recombination. These would normally be included as part of the Boltzmann collision operator for electron-neutral collisions, but the implementation used here has instead used effective rates from AMJUEL, so an effective cooling/heating operator needs to be implemented. The simplest implementation is a drag-like operator on $f_0$, which, given an inelastic electron-neutral process with energy cost $\epsilon$ and rate $K$, can be written as 

\begin{equation}
    \left(\frac{\partial f_0}{\partial t}\right)_{\text{inel}} = - \frac{K \epsilon n_n}{m_e v^2}\frac{\partial}{\partial v}\left(vf_0\right),
\end{equation}
which can be shown to reproduce the energy loss rate using partial integration on taking the second moment. However, a simple implementation will not preserve this analytical property, so the following velocity space discretization (in standard normalization) is used

\begin{equation}
    \left(\frac{\partial f_0}{\partial t}\right)_{\text{inel}} = - \frac{K \epsilon n_n}{v_k^2}\frac{C_kf_{0,k}-C_{k-1}f_{0,k-1}}{\Delta v_k},
\end{equation}
where now $C_k=v_k^2 \Delta v_k/(v_{k+1}^2-v_k^2)$ with boundary conditions $C_0=0$ and $C_{N_v}=\Delta v_{N_v}$. It is easy to show that this form gives the correct energy source. However, the number of particles is not necessarily conserved to machine precision due to the right boundary condition on $C_{N_v}$, which can be ensured by setting $C_{N_v}=0$, instead moving the error to the energy source. In the tests presented here this was not done since the spurious density source is proportional to the value of the distribution function in the last cell, which is essentially 0. Either way, if the distribution function tail is well resolved this error will never play a role. 

\section{Implicit BDE integrator with fixed-point iterations}
\label{appendix:BDE}

While the general approach for the Backward Difference Euler (BDE) integrator in ReMKiT1D borrows heavily from SOL-KiT\cite{Mijin2021}, in the interest of clarity, it is useful to have a summary of the method here, as well as how time step length is controlled and how the variable data is stored in the implicit vector passed to PETSc matrix solvers. 

Firstly, the variable placement in the implicit variable vector is done in the following way. Given a list of implicit variables $v_n$, which are either fluid (depend only on the spatial index) or distribution (depend on spatial, harmonic, and velocity space indices) variables, they are flattened and ordered in the implicit vector $F$ as follows

\begin{gather}
 \vec{F}
 =
  \begin{bmatrix}
   \vec{F}_1  \\
   \vec{F}_2 \\
   \vdots \\
   \vec{F}_{N_x}
   \end{bmatrix},
\end{gather}
where $F_i$ is the sub-vector corresponding to spatial index $i$ and is given by, for example, 

\begin{gather}
 \vec{F}_i
 =
  \begin{bmatrix}
   v_{1,i}  \\
   \vec{v}_{2,i}(h,v) \\
   \vdots \\
   v_{N,i}
   \end{bmatrix},
\end{gather}
$N$ is the total number of implicit variables, and where $\vec{v}_{2,i}$ is a distribution variable vector at spatial index $i$, which is further flattened first in the harmonic index $h$ and the velocity space index $v$

\begin{gather}
 \vec{v}_{2,i}(h,v)
 =
  \begin{bmatrix}
   v_{2,i,1,1}  \\
   v_{2,i,1,2} \\
   \vdots \\
   v_{2,i,1,N_v} \\
   v_{2,i,2,1} \\
   \vdots \\
   v_{2,i,N_h,N_v}
   \end{bmatrix},
\end{gather}
where $N_h$ and $N_v$ are the total number of distribution function harmonics and the number of velocity space cells.

Collating all individual term matrices one arrives at the global matrix equation

\begin{equation}
    \frac{d \vec{F}}{dt} = M(\vec{F})\cdot \vec{F},
\end{equation}
where now $M(\vec{F})$ is in general a non-linear global PETSc matrix. The implicit BDE method with fixed-point iterations then discretizes the solution in time as 

\begin{equation}
    \frac{\vec{F}^{i+1}-\vec{F}^{i+1}}{\Delta t_i} = M(\vec{F}^{i^*})\cdot \vec{F}^{i+1},
\end{equation}
where now $i^*$ represents the value at the previous non-linear iteration of the solver. The integrator converges based on a set of convergence variables and a relative or absolute non-linear iteration error\footnote{Usually based on $L_2$ norm of spatially-local values or largest $L_1$ norm}. Finally, the global (at the \textbf{CompositeIntegrator} level) time step length $\Delta t_i$ can be controlled through the application of a \textbf{TimestepController}, which scales the time step based on some spatially global criterion. An example is scaling an initial time step based on collisionality by multiplying it by the smallest value of (normalized) $T^{3/2}/n$, making sure that the shortest collisional times in the domain are always resolved. This is used in the SOL-KiT-like models in \ref{appendix:SOL-SK}.

Finally, the BDE integrator in ReMKiT1D is capable of recovering from failed matrix solves and cases where non-linear iterations fail by subdividing its time-step into smaller steps when a failure is detected. While crude, this method enables convergence when transient effects might momentarily make the matrix stiffer than expected. This integrator recovery is in addition to any time-step control defined at the global level. 



\bibliographystyle{elsarticle-num}
\bibliography{mybib.bib}







\end{document}